\documentclass[12pt,a4paper]{article}	
\usepackage{amsmath} 														%
\usepackage{amssymb}														%
\usepackage{color}															%

\usepackage{setspace}														%
\usepackage[left=3.75cm,right=3.75cm,top=3cm, bottom=3cm]{geometry}

\usepackage{graphicx}														%

\usepackage{url}																%
\usepackage{natbib}															%
\bibpunct{(}{)}{;}{a}{,}{,}											%

\usepackage[T1]{fontenc}

\usepackage[plainpages=false]{hyperref}					%

\usepackage{pst-all}                            %
\usepackage{booktabs}

\newcommand{\mI}{\mathcal{I}}
\newcommand{\mP}{\mathbb{P}}
\newcommand{\mF}{\mathbb{F}}
\newcommand{\mg}{\mathfrak{g}}

\newcommand{\fez}{\frac{1}{2}}

\newcommand{\oline}[1]{\overline{#1}}

\newtheorem{Proposition}{Proposition}
\newtheorem{Definition}{Definition}

\definecolor{snow}{rgb}{1.000,0.980,0.980}
\definecolor{snow1}{rgb}{1.000,0.980,0.980}
\definecolor{snow2}{rgb}{0.933,0.914,0.914}
\definecolor{snow3}{rgb}{0.804,0.788,0.788}
\definecolor{snow4}{rgb}{0.545,0.537,0.537}
\definecolor{GhostWhite}{rgb}{0.973,0.973,1.000}
\definecolor{WhiteSmoke}{rgb}{0.961,0.961,0.961}
\definecolor{gainsboro}{rgb}{0.863,0.863,0.863}
\definecolor{FloralWhite}{rgb}{1.000,0.980,0.941}
\definecolor{OldLace}{rgb}{0.992,0.961,0.902}
\definecolor{linen}{rgb}{0.980,0.941,0.902}
\definecolor{AntiqueWhite}{rgb}{0.980,0.922,0.843}
\definecolor{PapayaWhip}{rgb}{1.000,0.937,0.835}
\definecolor{BlanchedAlmond}{rgb}{1.000,0.922,0.804}
\definecolor{bisque}{rgb}{1.000,0.894,0.769}
\definecolor{PeachPuff}{rgb}{1.000,0.855,0.725}
\definecolor{NavajoWhite}{rgb}{1.000,0.871,0.678}
\definecolor{moccasin}{rgb}{1.000,0.894,0.710}
\definecolor{cornsilk}{rgb}{1.000,0.973,0.863}
\definecolor{ivory}{rgb}{1.000,1.000,0.941}
\definecolor{LemonChiffon}{rgb}{1.000,0.980,0.804}
\definecolor{seashell}{rgb}{1.000,0.961,0.933}
\definecolor{honeydew}{rgb}{0.941,1.000,0.941}
\definecolor{MintCream}{rgb}{0.961,1.000,0.980}
\definecolor{azure}{rgb}{0.941,1.000,1.000}
\definecolor{AliceBlue}{rgb}{0.941,0.973,1.000}
\definecolor{lavender}{rgb}{0.902,0.902,0.980}
\definecolor{LavenderBlush}{rgb}{1.000,0.941,0.961}
\definecolor{MistyRose}{rgb}{1.000,0.894,0.882}
\definecolor{white}{rgb}{1.000,1.000,1.000}
\definecolor{black}{rgb}{0.000,0.000,0.000}
\definecolor{DarkSlateGray}{rgb}{0.184,0.310,0.310}
\definecolor{DimGray}{rgb}{0.412,0.412,0.412}
\definecolor{SlateGray}{rgb}{0.439,0.502,0.565}
\definecolor{LightSlateGray}{rgb}{0.467,0.533,0.600}
\definecolor{gray}{rgb}{0.745,0.745,0.745}
\definecolor{LightGray}{rgb}{0.827,0.827,0.827}
\definecolor{MidnightBlue}{rgb}{0.098,0.098,0.439}
\definecolor{navy}{rgb}{0.000,0.000,0.502}
\definecolor{NavyBlue}{rgb}{0.000,0.000,0.502}
\definecolor{CornflowerBlue}{rgb}{0.392,0.584,0.929}
\definecolor{DarkSlateBlue}{rgb}{0.282,0.239,0.545}
\definecolor{SlateBlue}{rgb}{0.416,0.353,0.804}
\definecolor{MediumSlateBlue}{rgb}{0.482,0.408,0.933}
\definecolor{LightSlateBlue}{rgb}{0.518,0.439,1.000}
\definecolor{MediumBlue}{rgb}{0.000,0.000,0.804}
\definecolor{RoyalBlue}{rgb}{0.255,0.412,0.882}
\definecolor{blue}{rgb}{0.000,0.000,1.000}
\definecolor{DodgerBlue}{rgb}{0.118,0.565,1.000}
\definecolor{DeepSkyBlue}{rgb}{0.000,0.749,1.000}
\definecolor{SkyBlue}{rgb}{0.529,0.808,0.922}
\definecolor{LightSkyBlue}{rgb}{0.529,0.808,0.980}
\definecolor{SteelBlue}{rgb}{0.275,0.510,0.706}
\definecolor{LightSteelBlue}{rgb}{0.690,0.769,0.871}
\definecolor{LightBlue}{rgb}{0.678,0.847,0.902}
\definecolor{PowderBlue}{rgb}{0.690,0.878,0.902}
\definecolor{PaleTurquoise}{rgb}{0.686,0.933,0.933}
\definecolor{DarkTurquoise}{rgb}{0.000,0.808,0.820}
\definecolor{MediumTurquoise}{rgb}{0.282,0.820,0.800}
\definecolor{turquoise}{rgb}{0.251,0.878,0.816}
\definecolor{cyan}{rgb}{0.000,1.000,1.000}
\definecolor{LightCyan}{rgb}{0.878,1.000,1.000}
\definecolor{CadetBlue}{rgb}{0.373,0.620,0.627}
\definecolor{MediumAquamarine}{rgb}{0.400,0.804,0.667}
\definecolor{aquamarine}{rgb}{0.498,1.000,0.831}
\definecolor{DarkGreen}{rgb}{0.000,0.392,0.000}
\definecolor{DarkOliveGreen}{rgb}{0.333,0.420,0.184}
\definecolor{DarkSeaGreen}{rgb}{0.561,0.737,0.561}
\definecolor{SeaGreen}{rgb}{0.180,0.545,0.341}
\definecolor{MediumSeaGreen}{rgb}{0.235,0.702,0.443}
\definecolor{LightSeaGreen}{rgb}{0.125,0.698,0.667}
\definecolor{PaleGreen}{rgb}{0.596,0.984,0.596}
\definecolor{SpringGreen}{rgb}{0.000,1.000,0.498}
\definecolor{LawnGreen}{rgb}{0.486,0.988,0.000}
\definecolor{green}{rgb}{0.000,1.000,0.000}
\definecolor{chartreuse}{rgb}{0.498,1.000,0.000}
\definecolor{MediumSpringGreen}{rgb}{0.000,0.980,0.604}
\definecolor{GreenYellow}{rgb}{0.678,1.000,0.184}
\definecolor{LimeGreen}{rgb}{0.196,0.804,0.196}
\definecolor{YellowGreen}{rgb}{0.604,0.804,0.196}
\definecolor{ForestGreen}{rgb}{0.133,0.545,0.133}
\definecolor{OliveDrab}{rgb}{0.420,0.557,0.137}
\definecolor{DarkKhaki}{rgb}{0.741,0.718,0.420}
\definecolor{khaki}{rgb}{0.941,0.902,0.549}
\definecolor{PaleGoldenrod}{rgb}{0.933,0.910,0.667}
\definecolor{LightGoldenrodYellow}{rgb}{0.980,0.980,0.824}
\definecolor{LightYellow}{rgb}{1.000,1.000,0.878}
\definecolor{yellow}{rgb}{1.000,1.000,0.000}
\definecolor{gold}{rgb}{1.000,0.843,0.000}
\definecolor{LightGoldenrod}{rgb}{0.933,0.867,0.510}
\definecolor{goldenrod}{rgb}{0.855,0.647,0.125}
\definecolor{DarkGoldenrod}{rgb}{0.722,0.525,0.043}
\definecolor{RosyBrown}{rgb}{0.737,0.561,0.561}
\definecolor{IndianRed}{rgb}{0.804,0.361,0.361}
\definecolor{SaddleBrown}{rgb}{0.545,0.271,0.075}
\definecolor{sienna}{rgb}{0.627,0.322,0.176}
\definecolor{peru}{rgb}{0.804,0.522,0.247}
\definecolor{burlywood}{rgb}{0.871,0.722,0.529}
\definecolor{beige}{rgb}{0.961,0.961,0.863}
\definecolor{wheat}{rgb}{0.961,0.871,0.702}
\definecolor{SandyBrown}{rgb}{0.957,0.643,0.376}
\definecolor{tan}{rgb}{0.824,0.706,0.549}
\definecolor{chocolate}{rgb}{0.824,0.412,0.118}
\definecolor{firebrick}{rgb}{0.698,0.133,0.133}
\definecolor{brown}{rgb}{0.647,0.165,0.165}
\definecolor{DarkSalmon}{rgb}{0.914,0.588,0.478}
\definecolor{salmon}{rgb}{0.980,0.502,0.447}
\definecolor{LightSalmon}{rgb}{1.000,0.627,0.478}
\definecolor{orange}{rgb}{1.000,0.647,0.000}
\definecolor{DarkOrange}{rgb}{1.000,0.549,0.000}
\definecolor{coral}{rgb}{1.000,0.498,0.314}
\definecolor{LightCoral}{rgb}{0.941,0.502,0.502}
\definecolor{tomato}{rgb}{1.000,0.388,0.278}
\definecolor{OrangeRed}{rgb}{1.000,0.271,0.000}
\definecolor{red}{rgb}{1.000,0.000,0.000}
\definecolor{HotPink}{rgb}{1.000,0.412,0.706}
\definecolor{DeepPink}{rgb}{1.000,0.078,0.576}
\definecolor{pink}{rgb}{1.000,0.753,0.796}
\definecolor{LightPink}{rgb}{1.000,0.714,0.757}
\definecolor{PaleVioletRed}{rgb}{0.859,0.439,0.576}
\definecolor{maroon}{rgb}{0.690,0.188,0.376}
\definecolor{MediumVioletRed}{rgb}{0.780,0.082,0.522}
\definecolor{VioletRed}{rgb}{0.816,0.125,0.565}
\definecolor{magenta}{rgb}{1.000,0.000,1.000}
\definecolor{violet}{rgb}{0.933,0.510,0.933}
\definecolor{plum}{rgb}{0.867,0.627,0.867}
\definecolor{orchid}{rgb}{0.855,0.439,0.839}
\definecolor{MediumOrchid}{rgb}{0.729,0.333,0.827}
\definecolor{DarkOrchid}{rgb}{0.600,0.196,0.800}
\definecolor{DarkViolet}{rgb}{0.580,0.000,0.827}
\definecolor{BlueViolet}{rgb}{0.541,0.169,0.886}
\definecolor{purple}{rgb}{0.627,0.125,0.941}
\definecolor{MediumPurple}{rgb}{0.576,0.439,0.859}
\definecolor{thistle}{rgb}{0.847,0.749,0.847}
\definecolor{seashell1}{rgb}{1.000,0.961,0.933}
\definecolor{seashell2}{rgb}{0.933,0.898,0.871}
\definecolor{seashell3}{rgb}{0.804,0.773,0.749}
\definecolor{seashell4}{rgb}{0.545,0.525,0.510}
\definecolor{AntiqueWhite1}{rgb}{1.000,0.937,0.859}
\definecolor{AntiqueWhite2}{rgb}{0.933,0.875,0.800}
\definecolor{AntiqueWhite3}{rgb}{0.804,0.753,0.690}
\definecolor{AntiqueWhite4}{rgb}{0.545,0.514,0.471}
\definecolor{bisque1}{rgb}{1.000,0.894,0.769}
\definecolor{bisque2}{rgb}{0.933,0.835,0.718}
\definecolor{bisque3}{rgb}{0.804,0.718,0.620}
\definecolor{bisque4}{rgb}{0.545,0.490,0.420}
\definecolor{PeachPuff1}{rgb}{1.000,0.855,0.725}
\definecolor{PeachPuff2}{rgb}{0.933,0.796,0.678}
\definecolor{PeachPuff3}{rgb}{0.804,0.686,0.584}
\definecolor{PeachPuff4}{rgb}{0.545,0.467,0.396}
\definecolor{NavajoWhite1}{rgb}{1.000,0.871,0.678}
\definecolor{NavajoWhite2}{rgb}{0.933,0.812,0.631}
\definecolor{NavajoWhite3}{rgb}{0.804,0.702,0.545}
\definecolor{NavajoWhite4}{rgb}{0.545,0.475,0.369}
\definecolor{LemonChiffon1}{rgb}{1.000,0.980,0.804}
\definecolor{LemonChiffon2}{rgb}{0.933,0.914,0.749}
\definecolor{LemonChiffon3}{rgb}{0.804,0.788,0.647}
\definecolor{LemonChiffon4}{rgb}{0.545,0.537,0.439}
\definecolor{cornsilk1}{rgb}{1.000,0.973,0.863}
\definecolor{cornsilk2}{rgb}{0.933,0.910,0.804}
\definecolor{cornsilk3}{rgb}{0.804,0.784,0.694}
\definecolor{cornsilk4}{rgb}{0.545,0.533,0.471}
\definecolor{ivory1}{rgb}{1.000,1.000,0.941}
\definecolor{ivory2}{rgb}{0.933,0.933,0.878}
\definecolor{ivory3}{rgb}{0.804,0.804,0.757}
\definecolor{ivory4}{rgb}{0.545,0.545,0.514}
\definecolor{honeydew1}{rgb}{0.941,1.000,0.941}
\definecolor{honeydew2}{rgb}{0.878,0.933,0.878}
\definecolor{honeydew3}{rgb}{0.757,0.804,0.757}
\definecolor{honeydew4}{rgb}{0.514,0.545,0.514}
\definecolor{LavenderBlush1}{rgb}{1.000,0.941,0.961}
\definecolor{LavenderBlush2}{rgb}{0.933,0.878,0.898}
\definecolor{LavenderBlush3}{rgb}{0.804,0.757,0.773}
\definecolor{LavenderBlush4}{rgb}{0.545,0.514,0.525}
\definecolor{MistyRose1}{rgb}{1.000,0.894,0.882}
\definecolor{MistyRose2}{rgb}{0.933,0.835,0.824}
\definecolor{MistyRose3}{rgb}{0.804,0.718,0.710}
\definecolor{MistyRose4}{rgb}{0.545,0.490,0.482}
\definecolor{azure1}{rgb}{0.941,1.000,1.000}
\definecolor{azure2}{rgb}{0.878,0.933,0.933}
\definecolor{azure3}{rgb}{0.757,0.804,0.804}
\definecolor{azure4}{rgb}{0.514,0.545,0.545}
\definecolor{SlateBlue1}{rgb}{0.514,0.435,1.000}
\definecolor{SlateBlue2}{rgb}{0.478,0.404,0.933}
\definecolor{SlateBlue3}{rgb}{0.412,0.349,0.804}
\definecolor{SlateBlue4}{rgb}{0.278,0.235,0.545}
\definecolor{RoyalBlue1}{rgb}{0.282,0.463,1.000}
\definecolor{RoyalBlue2}{rgb}{0.263,0.431,0.933}
\definecolor{RoyalBlue3}{rgb}{0.227,0.373,0.804}
\definecolor{RoyalBlue4}{rgb}{0.153,0.251,0.545}
\definecolor{blue1}{rgb}{0.000,0.000,1.000}
\definecolor{blue2}{rgb}{0.000,0.000,0.933}
\definecolor{blue3}{rgb}{0.000,0.000,0.804}
\definecolor{blue4}{rgb}{0.000,0.000,0.545}
\definecolor{DodgerBlue1}{rgb}{0.118,0.565,1.000}
\definecolor{DodgerBlue2}{rgb}{0.110,0.525,0.933}
\definecolor{DodgerBlue3}{rgb}{0.094,0.455,0.804}
\definecolor{DodgerBlue4}{rgb}{0.063,0.306,0.545}
\definecolor{SteelBlue1}{rgb}{0.388,0.722,1.000}
\definecolor{SteelBlue2}{rgb}{0.361,0.675,0.933}
\definecolor{SteelBlue3}{rgb}{0.310,0.580,0.804}
\definecolor{SteelBlue4}{rgb}{0.212,0.392,0.545}
\definecolor{DeepSkyBlue1}{rgb}{0.000,0.749,1.000}
\definecolor{DeepSkyBlue2}{rgb}{0.000,0.698,0.933}
\definecolor{DeepSkyBlue3}{rgb}{0.000,0.604,0.804}
\definecolor{DeepSkyBlue4}{rgb}{0.000,0.408,0.545}
\definecolor{SkyBlue1}{rgb}{0.529,0.808,1.000}
\definecolor{SkyBlue2}{rgb}{0.494,0.753,0.933}
\definecolor{SkyBlue3}{rgb}{0.424,0.651,0.804}
\definecolor{SkyBlue4}{rgb}{0.290,0.439,0.545}
\definecolor{LightSkyBlue1}{rgb}{0.690,0.886,1.000}
\definecolor{LightSkyBlue2}{rgb}{0.643,0.827,0.933}
\definecolor{LightSkyBlue3}{rgb}{0.553,0.714,0.804}
\definecolor{LightSkyBlue4}{rgb}{0.376,0.482,0.545}
\definecolor{SlateGray1}{rgb}{0.776,0.886,1.000}
\definecolor{SlateGray2}{rgb}{0.725,0.827,0.933}
\definecolor{SlateGray3}{rgb}{0.624,0.714,0.804}
\definecolor{SlateGray4}{rgb}{0.424,0.482,0.545}
\definecolor{LightSteelBlue1}{rgb}{0.792,0.882,1.000}
\definecolor{LightSteelBlue2}{rgb}{0.737,0.824,0.933}
\definecolor{LightSteelBlue3}{rgb}{0.635,0.710,0.804}
\definecolor{LightSteelBlue4}{rgb}{0.431,0.482,0.545}
\definecolor{LightBlue1}{rgb}{0.749,0.937,1.000}
\definecolor{LightBlue2}{rgb}{0.698,0.875,0.933}
\definecolor{LightBlue3}{rgb}{0.604,0.753,0.804}
\definecolor{LightBlue4}{rgb}{0.408,0.514,0.545}
\definecolor{LightCyan1}{rgb}{0.878,1.000,1.000}
\definecolor{LightCyan2}{rgb}{0.820,0.933,0.933}
\definecolor{LightCyan3}{rgb}{0.706,0.804,0.804}
\definecolor{LightCyan4}{rgb}{0.478,0.545,0.545}
\definecolor{PaleTurquoise1}{rgb}{0.733,1.000,1.000}
\definecolor{PaleTurquoise2}{rgb}{0.682,0.933,0.933}
\definecolor{PaleTurquoise3}{rgb}{0.588,0.804,0.804}
\definecolor{PaleTurquoise4}{rgb}{0.400,0.545,0.545}
\definecolor{CadetBlue1}{rgb}{0.596,0.961,1.000}
\definecolor{CadetBlue2}{rgb}{0.557,0.898,0.933}
\definecolor{CadetBlue3}{rgb}{0.478,0.773,0.804}
\definecolor{CadetBlue4}{rgb}{0.325,0.525,0.545}
\definecolor{turquoise1}{rgb}{0.000,0.961,1.000}
\definecolor{turquoise2}{rgb}{0.000,0.898,0.933}
\definecolor{turquoise3}{rgb}{0.000,0.773,0.804}
\definecolor{turquoise4}{rgb}{0.000,0.525,0.545}
\definecolor{cyan1}{rgb}{0.000,1.000,1.000}
\definecolor{cyan2}{rgb}{0.000,0.933,0.933}
\definecolor{cyan3}{rgb}{0.000,0.804,0.804}
\definecolor{cyan4}{rgb}{0.000,0.545,0.545}
\definecolor{DarkSlateGray1}{rgb}{0.592,1.000,1.000}
\definecolor{DarkSlateGray2}{rgb}{0.553,0.933,0.933}
\definecolor{DarkSlateGray3}{rgb}{0.475,0.804,0.804}
\definecolor{DarkSlateGray4}{rgb}{0.322,0.545,0.545}
\definecolor{aquamarine1}{rgb}{0.498,1.000,0.831}
\definecolor{aquamarine2}{rgb}{0.463,0.933,0.776}
\definecolor{aquamarine3}{rgb}{0.400,0.804,0.667}
\definecolor{aquamarine4}{rgb}{0.271,0.545,0.455}
\definecolor{DarkSeaGreen1}{rgb}{0.757,1.000,0.757}
\definecolor{DarkSeaGreen2}{rgb}{0.706,0.933,0.706}
\definecolor{DarkSeaGreen3}{rgb}{0.608,0.804,0.608}
\definecolor{DarkSeaGreen4}{rgb}{0.412,0.545,0.412}
\definecolor{SeaGreen1}{rgb}{0.329,1.000,0.624}
\definecolor{SeaGreen2}{rgb}{0.306,0.933,0.580}
\definecolor{SeaGreen3}{rgb}{0.263,0.804,0.502}
\definecolor{SeaGreen4}{rgb}{0.180,0.545,0.341}
\definecolor{PaleGreen1}{rgb}{0.604,1.000,0.604}
\definecolor{PaleGreen2}{rgb}{0.565,0.933,0.565}
\definecolor{PaleGreen3}{rgb}{0.486,0.804,0.486}
\definecolor{PaleGreen4}{rgb}{0.329,0.545,0.329}
\definecolor{SpringGreen1}{rgb}{0.000,1.000,0.498}
\definecolor{SpringGreen2}{rgb}{0.000,0.933,0.463}
\definecolor{SpringGreen3}{rgb}{0.000,0.804,0.400}
\definecolor{SpringGreen4}{rgb}{0.000,0.545,0.271}
\definecolor{green1}{rgb}{0.000,1.000,0.000}
\definecolor{green2}{rgb}{0.000,0.933,0.000}
\definecolor{green3}{rgb}{0.000,0.804,0.000}
\definecolor{green4}{rgb}{0.000,0.545,0.000}
\definecolor{chartreuse1}{rgb}{0.498,1.000,0.000}
\definecolor{chartreuse2}{rgb}{0.463,0.933,0.000}
\definecolor{chartreuse3}{rgb}{0.400,0.804,0.000}
\definecolor{chartreuse4}{rgb}{0.271,0.545,0.000}
\definecolor{OliveDrab1}{rgb}{0.753,1.000,0.243}
\definecolor{OliveDrab2}{rgb}{0.702,0.933,0.227}
\definecolor{OliveDrab3}{rgb}{0.604,0.804,0.196}
\definecolor{OliveDrab4}{rgb}{0.412,0.545,0.133}
\definecolor{DarkOliveGreen1}{rgb}{0.792,1.000,0.439}
\definecolor{DarkOliveGreen2}{rgb}{0.737,0.933,0.408}
\definecolor{DarkOliveGreen3}{rgb}{0.635,0.804,0.353}
\definecolor{DarkOliveGreen4}{rgb}{0.431,0.545,0.239}
\definecolor{khaki1}{rgb}{1.000,0.965,0.561}
\definecolor{khaki2}{rgb}{0.933,0.902,0.522}
\definecolor{khaki3}{rgb}{0.804,0.776,0.451}
\definecolor{khaki4}{rgb}{0.545,0.525,0.306}
\definecolor{LightGoldenrod1}{rgb}{1.000,0.925,0.545}
\definecolor{LightGoldenrod2}{rgb}{0.933,0.863,0.510}
\definecolor{LightGoldenrod3}{rgb}{0.804,0.745,0.439}
\definecolor{LightGoldenrod4}{rgb}{0.545,0.506,0.298}
\definecolor{LightYellow1}{rgb}{1.000,1.000,0.878}
\definecolor{LightYellow2}{rgb}{0.933,0.933,0.820}
\definecolor{LightYellow3}{rgb}{0.804,0.804,0.706}
\definecolor{LightYellow4}{rgb}{0.545,0.545,0.478}
\definecolor{yellow1}{rgb}{1.000,1.000,0.000}
\definecolor{yellow2}{rgb}{0.933,0.933,0.000}
\definecolor{yellow3}{rgb}{0.804,0.804,0.000}
\definecolor{yellow4}{rgb}{0.545,0.545,0.000}
\definecolor{gold1}{rgb}{1.000,0.843,0.000}
\definecolor{gold2}{rgb}{0.933,0.788,0.000}
\definecolor{gold3}{rgb}{0.804,0.678,0.000}
\definecolor{gold4}{rgb}{0.545,0.459,0.000}
\definecolor{goldenrod1}{rgb}{1.000,0.757,0.145}
\definecolor{goldenrod2}{rgb}{0.933,0.706,0.133}
\definecolor{goldenrod3}{rgb}{0.804,0.608,0.114}
\definecolor{goldenrod4}{rgb}{0.545,0.412,0.078}
\definecolor{DarkGoldenrod1}{rgb}{1.000,0.725,0.059}
\definecolor{DarkGoldenrod2}{rgb}{0.933,0.678,0.055}
\definecolor{DarkGoldenrod3}{rgb}{0.804,0.584,0.047}
\definecolor{DarkGoldenrod4}{rgb}{0.545,0.396,0.031}
\definecolor{RosyBrown1}{rgb}{1.000,0.757,0.757}
\definecolor{RosyBrown2}{rgb}{0.933,0.706,0.706}
\definecolor{RosyBrown3}{rgb}{0.804,0.608,0.608}
\definecolor{RosyBrown4}{rgb}{0.545,0.412,0.412}
\definecolor{IndianRed1}{rgb}{1.000,0.416,0.416}
\definecolor{IndianRed2}{rgb}{0.933,0.388,0.388}
\definecolor{IndianRed3}{rgb}{0.804,0.333,0.333}
\definecolor{IndianRed4}{rgb}{0.545,0.227,0.227}
\definecolor{sienna1}{rgb}{1.000,0.510,0.278}
\definecolor{sienna2}{rgb}{0.933,0.475,0.259}
\definecolor{sienna3}{rgb}{0.804,0.408,0.224}
\definecolor{sienna4}{rgb}{0.545,0.278,0.149}
\definecolor{burlywood1}{rgb}{1.000,0.827,0.608}
\definecolor{burlywood2}{rgb}{0.933,0.773,0.569}
\definecolor{burlywood3}{rgb}{0.804,0.667,0.490}
\definecolor{burlywood4}{rgb}{0.545,0.451,0.333}
\definecolor{wheat1}{rgb}{1.000,0.906,0.729}
\definecolor{wheat2}{rgb}{0.933,0.847,0.682}
\definecolor{wheat3}{rgb}{0.804,0.729,0.588}
\definecolor{wheat4}{rgb}{0.545,0.494,0.400}
\definecolor{tan1}{rgb}{1.000,0.647,0.310}
\definecolor{tan2}{rgb}{0.933,0.604,0.286}
\definecolor{tan3}{rgb}{0.804,0.522,0.247}
\definecolor{tan4}{rgb}{0.545,0.353,0.169}
\definecolor{chocolate1}{rgb}{1.000,0.498,0.141}
\definecolor{chocolate2}{rgb}{0.933,0.463,0.129}
\definecolor{chocolate3}{rgb}{0.804,0.400,0.114}
\definecolor{chocolate4}{rgb}{0.545,0.271,0.075}
\definecolor{firebrick1}{rgb}{1.000,0.188,0.188}
\definecolor{firebrick2}{rgb}{0.933,0.173,0.173}
\definecolor{firebrick3}{rgb}{0.804,0.149,0.149}
\definecolor{firebrick4}{rgb}{0.545,0.102,0.102}
\definecolor{brown1}{rgb}{1.000,0.251,0.251}
\definecolor{brown2}{rgb}{0.933,0.231,0.231}
\definecolor{brown3}{rgb}{0.804,0.200,0.200}
\definecolor{brown4}{rgb}{0.545,0.137,0.137}
\definecolor{salmon1}{rgb}{1.000,0.549,0.412}
\definecolor{salmon2}{rgb}{0.933,0.510,0.384}
\definecolor{salmon3}{rgb}{0.804,0.439,0.329}
\definecolor{salmon4}{rgb}{0.545,0.298,0.224}
\definecolor{LightSalmon1}{rgb}{1.000,0.627,0.478}
\definecolor{LightSalmon2}{rgb}{0.933,0.584,0.447}
\definecolor{LightSalmon3}{rgb}{0.804,0.506,0.384}
\definecolor{LightSalmon4}{rgb}{0.545,0.341,0.259}
\definecolor{orange1}{rgb}{1.000,0.647,0.000}
\definecolor{orange2}{rgb}{0.933,0.604,0.000}
\definecolor{orange3}{rgb}{0.804,0.522,0.000}
\definecolor{orange4}{rgb}{0.545,0.353,0.000}
\definecolor{DarkOrange1}{rgb}{1.000,0.498,0.000}
\definecolor{DarkOrange2}{rgb}{0.933,0.463,0.000}
\definecolor{DarkOrange3}{rgb}{0.804,0.400,0.000}
\definecolor{DarkOrange4}{rgb}{0.545,0.271,0.000}
\definecolor{coral1}{rgb}{1.000,0.447,0.337}
\definecolor{coral2}{rgb}{0.933,0.416,0.314}
\definecolor{coral3}{rgb}{0.804,0.357,0.271}
\definecolor{coral4}{rgb}{0.545,0.243,0.184}
\definecolor{tomato1}{rgb}{1.000,0.388,0.278}
\definecolor{tomato2}{rgb}{0.933,0.361,0.259}
\definecolor{tomato3}{rgb}{0.804,0.310,0.224}
\definecolor{tomato4}{rgb}{0.545,0.212,0.149}
\definecolor{OrangeRed1}{rgb}{1.000,0.271,0.000}
\definecolor{OrangeRed2}{rgb}{0.933,0.251,0.000}
\definecolor{OrangeRed3}{rgb}{0.804,0.216,0.000}
\definecolor{OrangeRed4}{rgb}{0.545,0.145,0.000}
\definecolor{red1}{rgb}{1.000,0.000,0.000}
\definecolor{red2}{rgb}{0.933,0.000,0.000}
\definecolor{red3}{rgb}{0.804,0.000,0.000}
\definecolor{red4}{rgb}{0.545,0.000,0.000}
\definecolor{DeepPink1}{rgb}{1.000,0.078,0.576}
\definecolor{DeepPink2}{rgb}{0.933,0.071,0.537}
\definecolor{DeepPink3}{rgb}{0.804,0.063,0.463}
\definecolor{DeepPink4}{rgb}{0.545,0.039,0.314}
\definecolor{HotPink1}{rgb}{1.000,0.431,0.706}
\definecolor{HotPink2}{rgb}{0.933,0.416,0.655}
\definecolor{HotPink3}{rgb}{0.804,0.376,0.565}
\definecolor{HotPink4}{rgb}{0.545,0.227,0.384}
\definecolor{pink1}{rgb}{1.000,0.710,0.773}
\definecolor{pink2}{rgb}{0.933,0.663,0.722}
\definecolor{pink3}{rgb}{0.804,0.569,0.620}
\definecolor{pink4}{rgb}{0.545,0.388,0.424}
\definecolor{LightPink1}{rgb}{1.000,0.682,0.725}
\definecolor{LightPink2}{rgb}{0.933,0.635,0.678}
\definecolor{LightPink3}{rgb}{0.804,0.549,0.584}
\definecolor{LightPink4}{rgb}{0.545,0.373,0.396}
\definecolor{PaleVioletRed1}{rgb}{1.000,0.510,0.671}
\definecolor{PaleVioletRed2}{rgb}{0.933,0.475,0.624}
\definecolor{PaleVioletRed3}{rgb}{0.804,0.408,0.537}
\definecolor{PaleVioletRed4}{rgb}{0.545,0.278,0.365}
\definecolor{maroon1}{rgb}{1.000,0.204,0.702}
\definecolor{maroon2}{rgb}{0.933,0.188,0.655}
\definecolor{maroon3}{rgb}{0.804,0.161,0.565}
\definecolor{maroon4}{rgb}{0.545,0.110,0.384}
\definecolor{VioletRed1}{rgb}{1.000,0.243,0.588}
\definecolor{VioletRed2}{rgb}{0.933,0.227,0.549}
\definecolor{VioletRed3}{rgb}{0.804,0.196,0.471}
\definecolor{VioletRed4}{rgb}{0.545,0.133,0.322}
\definecolor{magenta1}{rgb}{1.000,0.000,1.000}
\definecolor{magenta2}{rgb}{0.933,0.000,0.933}
\definecolor{magenta3}{rgb}{0.804,0.000,0.804}
\definecolor{magenta4}{rgb}{0.545,0.000,0.545}
\definecolor{orchid1}{rgb}{1.000,0.514,0.980}
\definecolor{orchid2}{rgb}{0.933,0.478,0.914}
\definecolor{orchid3}{rgb}{0.804,0.412,0.788}
\definecolor{orchid4}{rgb}{0.545,0.278,0.537}
\definecolor{plum1}{rgb}{1.000,0.733,1.000}
\definecolor{plum2}{rgb}{0.933,0.682,0.933}
\definecolor{plum3}{rgb}{0.804,0.588,0.804}
\definecolor{plum4}{rgb}{0.545,0.400,0.545}
\definecolor{MediumOrchid1}{rgb}{0.878,0.400,1.000}
\definecolor{MediumOrchid2}{rgb}{0.820,0.373,0.933}
\definecolor{MediumOrchid3}{rgb}{0.706,0.322,0.804}
\definecolor{MediumOrchid4}{rgb}{0.478,0.216,0.545}
\definecolor{DarkOrchid1}{rgb}{0.749,0.243,1.000}
\definecolor{DarkOrchid2}{rgb}{0.698,0.227,0.933}
\definecolor{DarkOrchid3}{rgb}{0.604,0.196,0.804}
\definecolor{DarkOrchid4}{rgb}{0.408,0.133,0.545}
\definecolor{purple1}{rgb}{0.608,0.188,1.000}
\definecolor{purple2}{rgb}{0.569,0.173,0.933}
\definecolor{purple3}{rgb}{0.490,0.149,0.804}
\definecolor{purple4}{rgb}{0.333,0.102,0.545}
\definecolor{MediumPurple1}{rgb}{0.671,0.510,1.000}
\definecolor{MediumPurple2}{rgb}{0.624,0.475,0.933}
\definecolor{MediumPurple3}{rgb}{0.537,0.408,0.804}
\definecolor{MediumPurple4}{rgb}{0.365,0.278,0.545}
\definecolor{thistle1}{rgb}{1.000,0.882,1.000}
\definecolor{thistle2}{rgb}{0.933,0.824,0.933}
\definecolor{thistle3}{rgb}{0.804,0.710,0.804}
\definecolor{thistle4}{rgb}{0.545,0.482,0.545}
\definecolor{gray5}{rgb}{0.051,0.051,0.051}
\definecolor{gray10}{rgb}{0.102,0.102,0.102}
\definecolor{gray15}{rgb}{0.149,0.149,0.149}
\definecolor{gray20}{rgb}{0.200,0.200,0.200}
\definecolor{gray25}{rgb}{0.251,0.251,0.251}
\definecolor{gray30}{rgb}{0.302,0.302,0.302}
\definecolor{gray35}{rgb}{0.349,0.349,0.349}
\definecolor{gray40}{rgb}{0.400,0.400,0.400}
\definecolor{gray45}{rgb}{0.451,0.451,0.451}
\definecolor{gray50}{rgb}{0.498,0.498,0.498}
\definecolor{gray55}{rgb}{0.549,0.549,0.549}
\definecolor{gray60}{rgb}{0.600,0.600,0.600}
\definecolor{gray65}{rgb}{0.651,0.651,0.651}
\definecolor{gray70}{rgb}{0.702,0.702,0.702}
\definecolor{gray75}{rgb}{0.749,0.749,0.749}
\definecolor{gray80}{rgb}{0.800,0.800,0.800}
\definecolor{gray85}{rgb}{0.851,0.851,0.851}
\definecolor{gray90}{rgb}{0.898,0.898,0.898}
\definecolor{gray95}{rgb}{0.949,0.949,0.949}
\definecolor{gray100}{rgb}{1.000,1.000,1.000}
\definecolor{DarkGray}{rgb}{0.663,0.663,0.663}
\definecolor{DarkBlue}{rgb}{0.000,0.000,0.545}
\definecolor{DarkCyan}{rgb}{0.000,0.545,0.545}
\definecolor{DarkMagenta}{rgb}{0.545,0.000,0.545}
\definecolor{DarkRed}{rgb}{0.545,0.000,0.000}
\definecolor{LightGreen}{rgb}{0.565,0.933,0.565}

\psset{unit=1mm}
\begin{document}
\pagestyle{plain}
\setcounter{page}{1}

\title{Contagious Synchronization and Endogenous Network Formation in Financial Networks
\footnote{We would like to thank Toni Ahnert, Jean-Edouard Colliard, Jens Krause, Tarik Roukny, three anonymous referees, seminar participants at ECB, Bundesbank, as well as the 2013 INET Plenary Conference in Hong Kong, and the VIII Financial Stability Seminar organized by the Banco Central do Brazil for helpful discussions and comments. This paper has been prepared under the Lamfalussy Fellowship Program sponsored by the ECB whose support is gratefully ackonwledged. The views expressed in this paper do not necessarily reflect the views of Deutsche Bundesbank, the ECB, or the ESCB.}
}

\author{Christoph Aymanns\footnote{Mathematical Institute, University of Oxford and Institute for New Economic Thinking at the Oxford Martin School. E-Mail: \url{aymanns@maths.ox.ac.uk}.}, Co-Pierre Georg\footnote{Deutsche Bundesbank and University of Cape Town Graduate School of Business. E-Mail: \url{co-pierre.georg@bundesbank.de}.}}

\date{July 1, 2014}
\maketitle

\begin{abstract}
When banks choose similar investment strategies the financial system becomes vulnerable to common shocks. 
We model a simple financial system in which banks decide about their investment strategy based on a private belief about the state of the world and a social belief formed from observing the actions of peers.
Observing a larger group of peers conveys more information and thus leads to a stronger social belief.
Extending the standard model of Bayesian updating in social networks, we show that the probability that banks synchronize their investment strategy on a state non-matching action critically depends on the weighting between private and social belief.
This effect is alleviated when banks choose their peers endogenously in a network formation process, internalizing the externalities arising from social learning.
\par\medskip
{\bfseries Keywords:} social learning, endogenous financial networks, multi-agent simulations, systemic risk
\par\medskip
{\bfseries JEL Classification:} G21, C73, D53, D85
\end{abstract}
\clearpage
\section{Introduction}\label{Sec::Introduction}
When a large number of financial intermediaries choose the same investment strategy (i.e. their portfolios are very similar) the financial system as a whole becomes vulnerable to common shocks. A case at hand is the financial crisis of 2007/2008 when many banks invested into mortgage backed securities in anticipation that the underlying mortgages--many of which being US subprime mortgages--would not simultaneously depreciate in value. This assumption turned out to be incorrect resulting in one of the largest financial crises since the great depression. How could so many banks choose a non-optimal investment strategy despite the fact that they carefully monitor both economic fundamentals and the actions of other banks?\\

This paper presents a simple agent-based model in which financial intermediaries synchronize their investment strategy on a state non-matching action despite informative private signals about the state of the world. In a countable number of time-steps $N$ agents, representing financial intermediaries (banks for short), choose one of two actions. There are two states of the world which are revealed at the end of the simulation. A bank's action is either state-matching, in which case the bank receives a positive payoff if the state is revealed, or it is state-non-matching in which case the bank receives zero. Banks are connected to a set of peers in a financial network of mutual lines of credit resembling the interbank market. They receive a private signal about the state of the world and observe the previous actions of banks with whom they are connected via a mutual line of credit, but not of other banks. Based on both signals banks form a belief about the state of the world and choose their action accordingly.\\

Our model differs from the existing literature along two dimensions. First and foremost we develop an agent-based model of the financial system with strategic interaction amongst agents. This differs from existing models (see, for example, \cite{Poledna2014}, \cite{Bluhm2013}, \cite{Georg2013Interbank}, and \cite{Ladley2013}) where agent behaviour is myopic. Agents in myopic models react to the state of the world but when choosing an optimal action they do not take into account how other agents will react to their choice. Thus, the notion of equilibrium in myopic models is a rather mechanical. Strategic interaction amongst agents arises in our model from the fact that agents learn about their neighbors' actions, i.e. via the social belief. All the aforementioned papers furthermore use an exogenous network structure as starting point for the agent-based simulation, while our model uses an endogenous network formation process to arrive at a pairwise stable network structure that maximizes expected utility from social learning.\\

Second, while our model is mildly boundedly rational it shares a number of assumptions with the literature on Bayesian learning in social networks. The main difference to this literature (see, for example, \cite{Acemoglu2011}, \cite{GaleKariv2003}) is that we model an externality that is not present in the standard model of Bayesian learning in social networks. We assume that banks receive more information about the actions of other banks than they can computationally use. This assumption seems natural in a financial system that is increasingly complex.\footnote{See, for example, \cite{Haldane2012} for a discussion of increasing complexity in financial regulation.} The underlying assumption is that banks cannot adjust their actions (i.e. their investment strategy) as fast as they receive information from their peers and thus have to aggregate over potentially large amounts of information.\footnote{This assumption renders the agents boundedly rational, albeit mildly so, as for example \cite{DeMarzoVayanosZwiebel2003} argue.} The social belief in our model is formed not just from observing one neighbor at a time, but rather from observing a set of neighbors simultaneously. It is thus reasonable to assume that receipt of more information, i.e. observing the actions of a larger subset of agents, will create a stronger social belief than the receipt of less information. We model this by allowing for different weights of the social and private belief.\\

The other key difference to the existing literature on Bayesian learning in social networks is that we allow agents to endogenously form links based on the utility they get from an improved social belief in a first stage of the model. In the second stage of the model agents then learn about the state of the world and take their investment decisions. To the best of our knowledge, the only other paper considering endogenously formed social networks in a Bayesian learning setup is \cite{Acemoglu2014_Endogenous} who develop a two-stage game similar to ours. \cite{Acemoglu2014_Endogenous} model endogenous network formation via a communication cost matrix where some agents (in a {\it social clique}) can communicate at low costs, while others communicate at high cost. The main difference to our model is that we endogenously obtain a decreasing marginal value of additional links. We obtain a resulting endogenous network structure which is pairwise stable in the sense of \cite{JacksonWollinsky1996}.\\

We obtain two sets of results, one for networks with exogenous network structure, and one for endogenously formed networks. First, we analyze different ways of weighting private and social belief. In particular, we compare the standard {\it equal weighting} scenario in which agents place equal weights on their private and social belief, with two scenarios where agents place more weight on the social belief when they have a larger neighborhood. In the {\it neighborhood size} scenario the social belief is weighted with the size of the neighborhood, i.e. the private signal is weighted equal to every observed neighbor action. In the {\it relative neighborhood} scenario agents put more weight on the social belief when the neighborhood constitutes a larger share of the overall network. For completely uninformative signals there is no difference between these weighting functions. For informative signals, however, the weighting function has an impact on the probability that agents synchronize their investment decisions on a state non-matching action, i.e. for the probability that choosing a state-non-matching action is contagious.\footnote{Such {\it informational cascades} are a well-documented empirical phenomenon. See, for example, \cite{AlevyHaighList2006}, \cite{BernhardtCampelloKutsoati2006}, \cite{ChangChengKhorana2000}, \cite{ChiangZheng2010}, and \cite{CiprianiGuarino2014}.} Contagion is, very generally, understood as the transmission of adverse effects from one agent to another and is more likely if agents place greater weight on their social belief and depends on the density of the underlying exogenous network structure.\footnote{For a more thorough discussion of the different forms of contagion, see for example \cite{deBandtHartmannPeydro.2009}.} The probability of contagion increases by a factor of $200$ in the neighborhood size scenario compared to the equal weighting scenario, which highlights the importance of understanding the learning dynamics when agents place different weight on their social belief, depending on the size of their neighborhood.\\

We show that contagious synchronization occurs even if private signals are informative and if agents are initialized with an action that is on average state matching. The probability of contagion depends non-monotonously on the density of the network. For small network densities $\rho \lesssim 0.1$ the probability of contagion increases sharply and then decreases slowly for larger network densities. We confirm the robustness of our results by conducting $2,000$ independent simulations where we observe the average final action as a function of  the average initial action with varying network densities.\\

This result is of particular interest for policy makers as it relates two sources of systemic risk: common shocks and interbank market freezes. When the network density is too small, for example in the aftermath of an interbank market freeze, banks are unable to fully incorporate the information about their peers' actions. This effect is empirically tested by \cite{Caballero2012}, who documents a higher correlation amongst various asset classes in the world in the aftermath of the Lehman insolvency, i.e. during times of extreme stress on interbank markets and heightened uncertainty about the state of the world.
This can be understood as a contagious synchronization of bank's investment strategies for which our model provides a simple rationale.\\

Second, turning to the extension of endogenously formed networks, we show that endogenous link formation in the first stage of our model can significantly improve the speed of learning and reduce the probability of contagious synchronization relative to random networks. When private signals are less informative, the additional utility from forming a link is smaller and the endogenously formed network is less dense. This in turn can increase the probability of contagious synchronization in the second stage of the model. Heightened uncertainty about the state of the world, i.e. a less informative signal, does therefore not only directly increase the probability of contagious synchronization, but also indirectly because agents have less incentives to endogenously form links. If agents are heterogenous in the informativeness of their private signals, i.e. if some agents receive signals with higher precision than others, we show that the resulting endogenous network structure is of a core-periphery type. The structure of real-world interbank markets is often of this particular type, as for example \cite{CraigVonPeter2014} show. Naturally, these endogenously formed networks transfer information more effectively from highly informed agents to less informed agents than simple random networks.\\

This paper relates to three strands of literatures. First and foremost, the paper develops a financial multi-agent simulation in which agents learn not only from private signals, but also via endogenously formed interbank links. This is in contrast with existing multi-agent models of the financial system which include \cite{NierYangYorulmazerAlentorn.2007} and \cite{Iori2006} who take a fixed network and static balance sheet structure.\footnote{Closely related is the literature on financial networks. See, for example, \cite{AllenGale2000}, and \cite{FreixasParigiRochet.2000} for an early model of financial networks. The vast majority of models in this literature consider a fixed network structure only (see, amongst various others, \cite{Battiston2009}).} Slight deviations from these models can be found, for example, in \cite{Bluhm2013}, \cite{Ladley2013}, and \cite{Georg2013Interbank} who employ different equilibrium concepts. %
The main contribution this paper makes is to develop a sufficiently simple model of a financial system with a clear notion of equilibrium that allows to be implemented on a computer and tested against analytically tractable special cases.\\

Second, this paper relates to the literature on endogenous network formation pioneered by \cite{JacksonWollinsky1996} and \cite{BalaGoyal2000}. Few papers on endogenous network formation in interbank markets, exist, however. Notable exceptions are \cite{CastiglionesiNavarro2011} who study the formation of endogenous networks in a banking network with microfounded banking behaviour. Unlike \cite{CastiglionesiNavarro2011}, however, our paper uses a starkly simplified model of social learning to describe the behaviour of banks. This allows the introduction of informational spillovers from one bank to another, a mechanism not present in the work of \cite{CastiglionesiNavarro2011}.\\

Finally, our paper is closely related to the literature on Bayesian learning in social networks. The paper closest to ours in this literature is \cite{Acemoglu2014_Endogenous} who study a model of sequential learning in an endogenously formed social network where each agent receives a private signal about the state of the world and observe past actions of their neighbors. We contribute to this literature by allowing agents to place more weight on their social belief when their neighborhood is larger. Other related papers in this literature include \cite{Banerjee1992}, \cite{BikhchandaniHirshleiferWelch1992}, \cite{BalaGoyal1998}, and \cite{GaleKariv2003} who, however, all consider static networks only.\\

The remainder of this paper is organized as follows. The next section develops the baseline model and presents the results in the limiting case of an exogenous network structure. Section \ref{Sec::EndogenousNetwork} generalizes the model by allowing agents to endogenously form links in a first stage of the model. Section (\ref{Sec::Conclusion}) concludes.

\section{Contagious Synchronization with Fixed Network Structure}\label{Sec::ExogenousNetwork}
\subsection{Model Description and Timeline}\label{Sec::ExogenousNetwork:Model}
There is a countable number of dates $t=0,1,\ldots,T$ and a fixed number $i=1,\ldots,N$ of agents $A^i$ which represent financial institutions and are called banks for short. By a slight abuse of notation the model parameter $\theta$ is sometimes called the state of the world and we assume it can take two values $\theta\in\{0,1\}$. The probability that the world is in state $\theta$ is denoted as $\mP(\theta)$ and we assume that each state of the world is obtained with equal probability $\fez$. At each point in time $t$ bank $i$ chooses one of two investment strategies $x^i_t\in\{0,1\}$ which yields a positive return if the state of the world is revealed and matches the investment strategy chosen, and nothing otherwise. Agents take an action by choosing a certain investment strategy. Taking an action and switching between actions is costless. The utility of bank $i$ from investing is given as:
\begin{equation}\label{EQ::Utility}
 u^i(x^i,\theta) = \left\{ \begin{array}{ll}1&\textrm{ if } x^i=\theta\\0&\textrm{ else }\end{array}\right.
\end{equation}
The state of the world is unknown ex-ante and revealed at time $T$. This setup captures a situation where the state of the world is revealed less often (e.g. quarterly) than banks take investment decisions (e.g. daily).\footnote{In an alternative setup the state of the world is fixed throughout and an agent collects information and takes an irreversible decision at time $t$, but receives a payoff that is discounted by a factor $e^{-\kappa t}$. Both formulations incentivize agents to take a decision in finite time instead of collecting information until all uncertainty is eliminated.}\\

Banks can form interconnections in the form of mutual lines of credit. The set of banks is denoted $N=\{1,2,\ldots,n\}$ and the set of banks to which bank $i$ is directly connected is denoted $K^i\subseteq N$. Bank $i$ thus has $k^i = |K^i|$ direct connections called neighbors. This implements the notion of a network of banks $\mg$ which is defined as the set of banks together with a set of unordered pairs of banks called (undirected) links $L=\cup_{i=1}^n \{(i,j): j \in K^i\}$. A link is undirected since lines of credit are mutual and captured in the symmetric adjacency matrix $g$ of the network. Whenever a bank $i$ and $j$ have a link, the corresponding entry $g^{ij}=1$, otherwise $g^{ij}=0$. When there is no risk of confusion in notation, the network $\mg$ is identified by its adjacency matrix $g$. For the remainder of this section, I assume that the network structure is exogenously fixed and does not change over time. I assume that banks monitor each other continuously when granting a credit line and thus observe their respective actions.\\

In this section, the network $\mg$ is exogenously fixed throughout the simulation. In $t=0$ there is no previous decision of agents. Thus, each bank decides on its action in autarky. Banks receive a signal about the state of the world and form a private belief upon which they decide about their investment strategy $x^i_{t=0}$. The private signal received at time $t$ is denoted $s^i_t \in \oline{S}$ where $\oline{S}$ is a Euclidean space. Signals are independently generated according to a probability measure $\mF_\theta$ that depends on the state of the world $\theta$. The signal structure of the model is thus given by $(\mF_0,\mF_1)$. I assume that $\mF_0$ and $\mF_1$ are not identical and absolutely continuous with respect to each other. Throughout this paper I will assume that $\mF_0$ and $\mF_1$ represent Gaussian distributions with mean and standard deviation $(\mu_0,\sigma_0)$ and $(\mu_1,\sigma_1)$ respectively.\\

In $t=1, \ldots$ bank $i$ again receives a signal $s^i_{t}$ but now also observes the $t-1$ actions $x^j_{t-1}$ of its neighbors $j\in K^i$. The model outlined in this section is implemented in a multi-agent simulation where banks are the agents. Date $t=0$ in the model timeline is the initialization period. Subsequent dates $t=1,\ldots,T$ are the update steps which are repeated until the state of the world is being revealed in state $T$. Once the state is revealed, returns are realized and measured. In the simulation results discussed in Section \ref{Sec::ExogenousNetwork:Results} the state of the world was revealed at the end of the simulation after the system has reached a steady state in which agents do not change their actions any more.\footnote{In practice this is ensured by having many more update steps than it takes the system to reach a steady state.}\\

Banks form a private belief at time $t$ based on their privately observed signal $s^i_t$ and a social belief based on the observed actions $x^j_{t-1}$ their neighboring banks took in the previous period. The first time banks choose an action is a special case of the update step with no previous decisions being taken. The information set $I^i_t$ of a bank $i$ at time $t$ is given by the private signal $s^i_t$, the set of banks connected to bank $i$ in $t-1$, $K^i_{t-1}$, and the actions $x^j_{t-1}$ of connected banks $j \in K^i_{t-1}$. Formally:
\begin{equation}\label{EQ::InformationSet}
 I^i_t = \left\{s^i_t, K^i_{t-1}, x^j_{t-1} \forall j \in K^i_{t-1}\right\}
\end{equation}
The set of all possible information sets of bank $i$ is denoted by $\mI^i$. A strategy for bank $i$ selects an action for each possible information set. Formally, a strategy for bank $i$ is a mapping $\sigma^i: \mI^i \rightarrow x^i = \left\{0,1\right\}$. The notation $\sigma^{-i} = \{\sigma^1,\ldots,\sigma^{i-1},\sigma^{i+1},\sigma^n\}$ is used to denote the strategies of all banks other than $i$.\\

A strategy profile $\sigma=\{\sigma^i\}_{i\in1,\ldots,n}$ is a pure strategy equilibrium of this game of social learning for a bank $i$'s investment, if $\sigma^i$ maximizes the bank's expected pay-off, given the strategies of all other banks $\sigma^{-i}$.
\cite{Acemoglu2011} show that the strategy decision of bank $i$, $x^i_t=\sigma^i(I^i_t)$ is given as:
\begin{equation}\label{EQ1}
   x^i = \begin{cases}
                  1 &\mbox{if } \mP_\sigma(\theta=1|s^i_t) + \mP_\sigma(\theta=1|x^j_{t-1}, j\in K^i_{t-1}) > 1\\
                  0 &\mbox{if } \underbrace{\mP_\sigma(\theta=1|s^i_t)}_{\textrm{private belief }p} + \underbrace{\mP_\sigma(\theta=1|x^j_{t-1}, j\in K^i_{t-1})}_{\textrm{social belief }q} < 1
         \end{cases}
\end{equation}
and $x^i\in\{0,1\}$ otherwise. The first term on the right-hand side of Equation \ref{EQ1} is the private belief, the second term is the social belief, and the threshold is fixed to $\fez$. This equation can be generalized when introducing weights on the private and social belief. In its most general form, it can be written as:
\begin{equation}\label{Eqn::Choice}
x^i = \begin{cases} 1 &\mbox{if } t(p^i,q^i) > \fez  \\
0 & \mbox{if } t(p^i,q^i) < \fez \end{cases}
\end{equation}
where $t(p^i,q^i)$ is a weighting function depending on the private and social belief. A simple weighting function 
\begin{equation}\label{Eqn::simpleWeight}
t(p^i,q^i) = \begin{cases} \fez ( p^i + q^i ) &\mbox{if } |K^i| > 0,  \\
p^i & |K^i| = 0, \end{cases}
\end{equation}
implements the model of \cite{Acemoglu2011} where agents place equal weight on their private and social belief (we denote this weighting function as the {\it equal weighting} scenario).\\

The simple weighting function is appropriate in a setting where agents receive two signals at a time only: their private signal and the signal of their direct predecessor in the social network. In our setting, however, banks receive multiple signals, only one of which is their private signal. It is thus natural to allow for more general weighting functions which, however, have to satisfy two conditions: (i) In the case with no social learning ($K^i_{t-1} = \emptyset$), the weighting should reduce to the simple case $t(p^i,q^i) = p^i$ in which the agent will select action $x^i = 1$ whenever it is more likely that the state of the world is $\theta=1$ and zero otherwise; and (ii) With social learning the weighting should depend on the number of neighboring signals, i.e. the size of the neighborhood $k^i_{t-1} = |K^i_{t-1}|$. The underlying assumption is that the agent will place a higher weight on the social belief when the neighborhood is larger. We consider two different scenarios for the weighting function: (i) The private signal and each observed action are equally weighted (called the {\it neighborhood size} scenario):
\begin{equation}
 t(p^i,q^i) = \left(\frac{1}{k^i_{t-1}+1}\right)p^i + \left(\frac{k^i_{t-1}}{k^i_{t-1}+1}\right)q^i
\end{equation}
And (ii) Observed actions are weighted with the relative size of the neighborhood (called the {\it relative neighborhood} scenario):
\begin{equation}
 t(p^i,q^i) = \left( 1 - \frac{k^i_{t-1}}{N-1}\right)p^i + \left(\frac{k^i_{t-1}}{N-1}\right)q^i
\end{equation}

The private belief of bank $i$ is denoted $p^i = \mP(\theta=1|s^i)$ and can easily be obtained using Bayes' rule. It is given as:
\begin{equation}\label{EQ:PrivateBelief}
 p^i = \left(1 + \frac{d\mF_0}{d\mF_1}(s^i_t)\right)^{-1} = \left(1 + \frac{f_0(s^i_t)}{f_1(s^i_t)}\right)^{-1}
\end{equation}
where $f_0$ and $f_1$ are the densities of $\mF_0$ and $\mF_1$ respectively. Bank $i$ is assumed to form a social belief $q^i$ by simply averaging over the actions of all neighbors $j\in K^i_{t-1}$:
\begin{equation}\label{EQ:SocialBelief}
 q^i = \mP_\sigma(\theta=1|K^i_t,x^j,j\in K^i_{t-1}) = 1/k^i_{t-1} \sum_{j\in K^i_{t-1}} x^j_{t-1}
\end{equation}
Given these private and social beliefs, agents choose an action according to equation (\ref{Eqn::Choice}).\\

Averaging over the actions of neighbors is a special case of \cite{DeGroot1974} who introduces a model where a population of $N$ agents is endowed with initial opinions $p(0)$. Agents are connected to each other but with varying levels of trust, i.e. their interconnectedness is captured in a weighted directed $n\times n$ matrix $T$. A vector of beliefs $p$ is updated such that $p(t) = T p(t-1) = T^t p(0)$. \cite{DeMarzoVayanosZwiebel2003} point out that this process is a boundedly rational approximation of a much more complicated inference problem where agents keep track of each bit of information to avoid a persuasion bias (effectively double-counting the same piece of information). Therefore, the model this paper develops is also boundedly rational.\footnote{This bounded rationality can be motivated analogously to \cite{DeMarzoVayanosZwiebel2003} who argue that the amount of information agents have to keep track of increases exponentially with the number of agents and increasing time, making it computationally impossible to process all available information.}\\

The model in this section can be formulated as an agent-based model. Banks are agents $a^i$ who choose one of two actions $x^i\in\{0,1\}$. Variables determined in the model internally are given by the private and social belief of agent $i$ at time $t$, $p^i_t,q^i_t$ and the only exogenously given parameter is the state of the world $\theta$ which is identical for all agents $i$. Each agent has an information set $I^i$ given by Equation (\ref{EQ::InformationSet}). Agents receive utility (\ref{EQ::Utility}) and decide on their optimal strategy given in Equation (\ref{EQ1}). The interaction of agents is captured in a network structure $\mg$, encapsulated in an agent $i$'s information set. Equations (\ref{EQ:PrivateBelief}) and (\ref{EQ:SocialBelief}) specify how agents take their decisions and choose an optimal strategy.

\subsection{Herding with Exogenous Network Structures}\label{Sec::ExogenousNetwork:Results}
Our interest is to understand under which conditions agents in the model with an exogenously fixed network structure coordinate on a state non-matching action. Before analyzing the full model, we build some intuition by discussing useful benchmark cases. Let the state of the world be $\theta=0$ and assume that $f_0 = m f_1$. For $m=1$ the signal is completely uninformative. For $m>1$ the signal is informative and more so the larger $m$ is. In the equal weighting scenario, equation (\ref{EQ1}) together with equations (\ref{EQ:PrivateBelief}) and (\ref{EQ:SocialBelief}) yields:
\begin{equation}\label{EQ:Equal1}
 \fez (1 + m)^{-1} + \fez \frac{1}{k} \sum_{j\in K^i_{t-1}} x^j_{t-1} > \fez \Leftrightarrow \sum_{j\in K^i_{t-1}} x^j_{t-1} > k\left[1-\frac{1}{(1+m)}\right]
\end{equation}
For completely uninformative signals, $m=1$, equation (\ref{EQ:Equal1}) implies that an agent will always follow the majority of her neighbors. For very informative signals, $m\gg 1$, equation (\ref{EQ:Equal1}) implies that an agent will only ignore her private signal if she receives a strong social signal. The required strength of the social signal increases with the precision of the private signal. Now consider the neighborhood size scenario. The equation analogous to (\ref{EQ:Equal1}) for this scenario reads:
\begin{eqnarray}\label{EQ:Neighbor1}
 \left(\frac{1}{k^i_{t-1} + 1}\right)(1 + m)^{-1} + \left(\frac{k^i_{t-1}}{k^i_{t-1}+1}\right) \frac{1}{k^i_{t-1}} \sum_{j\in K^i_{t-1}} x^j_{t-1} > \fez \\
 \quad \Leftrightarrow  \quad \sum_{j\in K^i_{t-1}} x^j_{t-1} > \frac{k^i_{t-1}}{2} + \left[ \fez - \frac{1}{1+m} \right]
\end{eqnarray}
For completely uninformative signals $m=1$ this condition reduces to the equal weighting scenario. For highly informative signals, $m \gg 1$, however, the agent is almost as willing to follow her neighbors as in the uninformative equal weighting scenario. The neighborhood scenario thus captures the situation where the agent is aware not only of her own private signal informativeness, but also of that of her neighbors. In the relative neighborhood scenario equation (\ref{EQ:Equal1}) reads:
\begin{eqnarray}
 \left(1 - \frac{k^i_{t-1}}{N-1} \right)(1+m)^{-1} + \left(\frac{k^i_{t-1}}{N-1}\right) \frac{1}{k^i_{t-1}} \sum_{j\in K^i_{t-1}} x^j_{t-1} > \fez \\
 \quad \Leftrightarrow \quad \sum_{j\in K^i_{t-1}} x^j_{t-1} > (N-1)\left[\fez - \frac{1}{1+m}\right] + \frac{k^i_{t-1}}{1+m}
\end{eqnarray}
Again, for completely uninformative signals this equation reduces to the equal weighting case. For highly informative signals, $m\gg 1$, this equation reduces to $\sum_{j\in K^i_{t-1}} x^j_{t-1} > \fez(N-1)$ if the neighborhood is sufficiently small $m \gg k^i_{t-1}$. An agent will thus ignore her private signal and follow the majority of her neighbors, if her neighborhood is larger than half of the network. The central node in a star network will thus always follow the majority of the spokes and spokes will always follow their private signal.\\

We can gain further insights into the model dynamics by resorting to a mean-field approximation in which we consider the simplified action dynamics of a representative agent. Given the adjacency matrix $g$ of a network $\mg$, the social belief $q$ of the representative agent is given as $q = g \mathbf{x} / k$ where $\mathbf{x}$ is the vector of all agent's actions. The social belief in the mean-field approximation is simply the average action of the population:
\begin{equation}
	q = \Pr(x = 0 \mid q = q)\cdot 0 + \Pr(x = 1 \mid q = q)\cdot 1 = \Pr(x = 1 \mid q = q)
\end{equation}
which yields a self-consistency relation for the social belief and hence for the average action of the population. The equilibrium average action $q^*$ is implicitely given by the solution to the self-consistency condition:
\begin{equation}
q^* = \Pr(x = 1 \mid q = q^*) = \int_{1-q*}^1 p_P(p \mid \theta = 0) dp.
\end{equation}
Note that  $\Pr(x = 1 \mid q = 0) = 0$ and $\Pr(x = 1 \mid q = 1) = 1$. Since $\Pr(x = 1 \mid q = q^*)$ is the cumulative distribution function of the bell shaped private belief it will have a sigmoid shape. Therefore the self consistency equation will have three solutions: $q_1=0$, $q_2=1$ and some $q_3 \in (0,1)$. We illustrate this in figure \ref{Fig::EquilibriumSocialBelief}. $q_1$, $q_2$ are stable fixed points while $q_3$ is unstable (this can be seen graphically in figure \ref{Fig::EquilibriumSocialBelief} and is a direct result from the specified learning dynamics). $q_3$ defines the ``critical'' social belief beyond which the system synchronizes on the state non-matching action.\\

These exercises show that the impact of the weighting function on agents' strategies is easily understood in the case of completely uninformative and fully informative signals or in a mean field approximation. But what happens in the more realistic interim region? Are densely connected networks more conducive for agents to coordinate on state non-matching actions or sparse networks? And how does the fraction of nodes that coordinate on a state non-matching action depend on the initial conditions? We address these questions in an agent-based simulation for three cases. In the case (I) of {\it informed agents} the distance between the mean of the two signals $\mu_1 - \mu_0 = 0.6-0.4 = 0.2$ while in the case (U) of uninformed agents the distance between the mean of the two signals is $\mu_1 - \mu_0 = 0.51 - 0.49 = 0.02$. In both cases we use a standard deviation of $\sigma_{0,1} = \sqrt{0.1}$.\footnote{A larger standard deviation would make the signal less informative, without changing our results qualitatively.} We conduct our simulations with $N=100$ agents and update $T=100$ times. To analyze the impact of the network structure on the probability of coordination on a state non-matching action, we vary the network density $\rho$ of a random (Erd\"os-R\'enyi) graph within $\rho=[0.0,0.95]$ in 20 steps. Each simulation is repeated $S=1,000$ times to account for stochasticity. For all simulations we assume that the state of the world is $\theta = 0$. An overview of the parameters used can be found in Table \ref{Table::OverviewParamters}.\\

Figure (\ref{Figure::ExoAvgFinalAction}) shows the average final action of the system after $T=100$ update steps for a random graph with varying densities in the informed and uninformed case for the equal weighting, neighborhood size, and relative neighborhood scenario. When the network density is very low, agents effectively act on the basis of their private signal only and the fraction of agents that choose a state non-matching action is proportional to the signal informativeness. With increasing network density, social learning sets in and the fraction of agents with a state non-matching action is reduced. With an informative signal and equal weighting there is almost no agent that chooses a state non-matching action after $T=100$ update steps. In the neighborhod size scenario agents are more likely to follow their neighbors than in the equal weighting scenario. The initial conditions of the simulation thus have a stronger effect on an agent's decision. It will therefore take a longer time for agents to choose a state matching action. As long as the signal is not fully informative, equation (\ref{EQ:Neighbor1}) implies that an agent $i$ will only choose action $x^i=1$ if more than half of her neighbors chose that action. With increasing network density  the dependence on the initial conditions will become more important, which explains the increase in the fraction of agents choosing a state non-matching action with increasing network density in the center-left panel of Figure (\ref{Figure::ExoAvgFinalAction}). Note that this effect is not present for uninformative signals as can be seen in the right panel of Figure (\ref{Figure::ExoAvgFinalAction}). Finally, agent $i$ in the relative neighborhood scenario chooses $x^i=1$ only if a relatively large fraction of her neighbors also chooses this action. The threshold for this is independent of the size of the neighborhood and only depends on the total network size.\\

The initial conditions become more important in the case of informative signals for the neighborhood size scenario. In order to understand how exactly our results depend on the initial conditions, Figure (\ref{Figure::ExoFracFinalAction}) shows the probability that a large fraction ($>80\%$) of agents coordinate on a state non-matching action for three cases: (1) For a full sample of $S=1,000$ simulations; (2) conditional on agents starting on average with a state matching action: $\hat{x} = \sum_i x^i(0)/N < \fez$; (3) conditional on agents starting on average with a state non-matching action: $\hat{x} > \fez$. As expected, the probability that agents coordinate on a state non-matching action drastically increases when agents initially start with a state non-matching action. However, less so in the equal weighting scenario because agents place less weight on their social beliefs and private signals are informative. A comparison of the left-center with the left-top panel in Figure (\ref{Figure::ExoFracFinalAction}) shows that the probability of contagion increases by a factor of roughly $200$ in the neighborhood size scenario compared to the equal weighting scenario. Again, for uninformative signals this effect is not present.\\

While Figure (\ref{Figure::ExoFracFinalAction}) shows the existence of contagion even for initial actions that are state matching, the relationship between average initial and average final action is not yet quantified. Therefore, in Figure (\ref{Figure::ExoAvgFinalVSAvgInitial}) we plot the average final action versus the average initial action in a density plot. We conducted a total of $S=1,000 \times 20$ simulations and show the resulting pair of average initial and average final action as a dot with the respective coordinates. The left side of Figure (\ref{Figure::ExoAvgFinalVSAvgInitial}) is for an informative signal and we draw initial actions according to the (informative) private signal. The mean of initial distributions is thus $\hat{x}_I < 0.5$, i.e. informative on average. In the right panel we show the same results for an uninformative signal and the mean of initial distributions is thus much closer to $\hat{x}_I = \fez$. Contagion is shown in the upper right quadrant of each subfigure. For the equal weighting scenario (top), only very few simulations yield a final average action that is state non-matching. A similar picture can be seen for the relative neighborhood (bottom) scenario. In the neighborhood size (center) scenario, however, a substantial number of simulations with an initially state matching average action yield a final state non-matching average action, confirming the existence of a contagious regime.
\section{Contagious Synchronization in Endogenously Formed Networks}\label{Sec::EndogenousNetwork}
Banks form interbank networks endogenously. The decision whether or not two banks engage in interbank lending, e.g. in the form of agreeing on a mutual line of credit, depends in reality on many factors, including liquidity needs and counterparty risk. In the previous section we analyzed how one bank can learn about an underlying state of the world by observing the action of another bank to which it has issued an interbank credit. This additional information can create a benefit for the loan-issuing bank that constitutes a, possibly positive, externality for the lending decision. When banks coordinate on a state non-matching information, however, ``learning'' about a neighboring bank's action constitutes a negative externality. The net effect of both externalities determines whether two banks are willing to engage in interbank lending. We compute the value of an additional link in three steps. First, we compute the probability that an agent chooses a state matching action, given her signal structure, private beliefs and neighbors' actions. Given this probability, we compute, second, an agent's expected utility conditional on her social belief which depends on her strategic choice to establish a link. Once an agent's expected utility with and without a link is computed, we can use the concept of pairwise stable networks to determine the equilibrium network structure.

\subsection{The Probability that Agents Choose a State Matching Action}\label{Sec::EndogenousNetwork:ProbStateMatchingAction}
Based on the signal structure, the update of private beliefs, and neighbors' action choice we can derive the probability that an agent $i$ chooses a state matching action. To see this, we first derive the distribution of private beliefs. The private signal structure is given as:
\begin{equation}\label{EQ::Endo:1}
	\begin{aligned}
		f_0(s) &= \frac{1}{\sqrt{2\pi}\sigma_0} \exp \left( \frac{-(s-\mu_0)^2}{2 \sigma_0^2} \right)\\
		f_1(s) &= \frac{1}{\sqrt{2\pi}\sigma_1} \exp \left( \frac{-(s-\mu_1)^2}{2 \sigma_1^2} \right)
	\end{aligned}
\end{equation}
and we assume that $\sigma_0 = \sigma_1 = \sigma$. Denote the probability distribution of agent $i$'s private belief $p^i$ as $f_p(p^i)$. We can then state the following:
\begin{Proposition}\label{Prop::Prop1}
For $\theta \in \{0,1\}$ the probability that agent $i$ chooses a state matching action $x^i=\theta$, given a social belief $q^i=q$ and private belief $p^i$, is given by:
\begin{equation}\label{EQ::Endo:2}
	\Pr(x^i = \theta \mid q^i = q) = \int_0^{1-q} f_p(p^i \mid \theta = 0) dp^i  =\int_{1-q}^1 f_p(p^i \mid \theta = 1) dp^i,
\end{equation}
where the distribution $f_p(p^i \mid \theta = 0)$ of agent $i$'s private belief is given as:
\begin{equation}
	f_p(p^i \mid \theta = 0) = \frac{\left(\left(-\frac{1-p}{p^2}-\frac{1}{p}\right) p \sigma^2\right) \exp \left(-\frac{\left((\mu_0-\mu_1)^2-2 \sigma^2 \log \left(\frac{1}{p}-1\right)\right)^2}{8 \sigma^2 (\mu_0-\mu_1)^2}\right)}{\left(\sqrt{2 \pi } \sigma\right) ((1-p) (\mu_0-\mu_1))}.
\end{equation}
and similarly for $f_p(p^i \mid \theta = 1)$.
\end{Proposition}
Proof, see Appendix (\ref{Appendix::Proofs}). \hfill  $\blacksquare$\\

In the absence of maturity, liquidity and counterparty risk, the value of an interbank loan, is proportional to the probability that the newly connected neighbor chooses a state-matching action and thus proportional to (\ref{EQ::Endo:2}).

\subsection{Agents' Expected Utility}\label{Sec::EndogenousNetwork:ExpectedUtility}
Recall that an agent $i$'s utility $u^i$ is given as:
\begin{equation}
u^i(x^i) = \begin{cases} 1 &\mbox{if } x^i = \theta  \\
0 & \mbox{if } x^i \neq \theta\end{cases}
\end{equation}
The expected utility of agent $i$ conditional on her social belief $q^i$ is thus:
\begin{equation}
	\oline{u}^i(x^i \mid q^i) = \Pr(x^i = \theta \mid q^i = q).
\end{equation}
The value of a link is given by the marginal utility from establishing a link, which in turn depends on the change in the social belief $q$. An agent can thus influence her social belief by strategically choosing neighbors. The probability that a neighbor takes a state matching action depends in turn on the social belief that this neighbor forms about her neighbors, which leads to complex higher-order effects which we neglect in this paper. Rather, we assume that an agent $i$ has constant beliefs about her neighbors' social beliefs $q'$. An agent who ignores the effect of second-nearest neighbors (i.e. neighbors of neighbors) on the social beliefs of nearest neighbors will simply assume that her neigbhors' social belief is $q'=\fez$. This amounts to assuming that the neighbors' actions are independent of each other. The expected utility of agent $i$ conditional on a given $q'$ and neighborhood $K^i$ is given as:
\begin{equation}\label{EQ::Endo:4}
 \oline{u}^i(q',K^i) = \sum_{a \in Q^i} \Pr(q^i = a \mid q') \Pr(x^i = \theta \mid q^i = a),
\end{equation}
where $Q^i$ is the set of all possible values of the social belief of agent $i$. The first term on the right-hand side of equation (\ref{EQ::Endo:4}) is the probability that agent $i$ has a certain social belief given the social belief of her neighbors. The second term is the probability of choosing a state matching action given that social belief and given by equation (\ref{EQ::Endo:2}). For a given size of the agent's neighborhood $k^i$, $Q^i$ is simply $Q^i = \{n/k^i \mid n \in \mathbb{Z}, 0 \leq n \leq k^i \}$. The probability of a particular social belief can be computed by summing over the probabilities of combinations of actions chosen by the neighbors of agent $i$. Define the set of feasible action vectors of $i$'s neighbors conditional on agent $i$ having a social belief $q^i=a$: 
\begin{equation}
	X^{ai} = \{\mathbf{x} \mid \sum_j x_j = a, x_j \in \{0,1\}, j\in K_i\}. 
\end{equation}
i.e. $X^{ai}$ is the set of all action vectors that are compatible with a social belief $q^i=a$. Then the probability of agent $i$ having a private belief of $q^i=a$ given the social beliefs of all $i$'s neighbors, $q'$, is given as:
\begin{equation}\label{EQ::Endo:3}
	\Pr(q^i = a \mid q') = \sum_{\mathbf{y} \in  X^{ai}} \prod_{j \in K^i} \Pr(x^j = y^j \mid q^j = q').
\end{equation}
Now define the probability that neighbor $j$ chooses a state matching action as:
\begin{equation}
z^j = \Pr(x^j = \theta \mid q^j = q') = \int_0^{1-q^j} f_p(p^j \mid \theta = 0) dp
\end{equation}
Note, that $f_p(p^j\mid \theta=0)$ depends on the signal structure of neighbor $j$.\\

We can write for the probability of agent $i$ having a private belief of $q^i=a$ in equation (\ref{EQ::Endo:3}):
\begin{equation}
	\Pr(x^j = y^j \mid q^j = q') = \begin{cases} z^j &\mbox{if } y^j = 0  \\
	1-z^j & \mbox{if } y^j = 1 \end{cases}
\end{equation}
If $z^j = z ~\forall j$ the distribution in \ref{EQ::Endo:3} would be a simple binomial distribution. In general, however, this is not the case and we need to resort to numerical methods to compute the equilibrium network structures.

\subsection{The Network Formation Process}\label{Sec::EndogenousNetwork:NetworkFormationProcess}
We now have all necessary ingredients to compute the expected utility of an agent $i$ given her neighborhood $K^i$ and expectations about her neighbors' social belief $q'$. In the endogenous network formation process the agent will seek to maximize her utility by changing her neighborhood while holding $q'$ fixed. In this section we outline an algorithm for endogenous network formation that ensures a pairwise stable network in the sense of \cite{JacksonWollinsky1996}:
\begin{Definition}
 A network defined by an adjacency matrix $g$ is called pairwise stable if
 \begin{enumerate}
  \item[(i)] For all banks $i$ and $j$ directly connected by a link, $l^{ij}\in L$: $u^i(g) \geq u^i(g-l^{ij})$ and $u^j(g) \geq u^j(g-l^{ij})$
  \item[(ii)] For all banks $i$ and $j$ not directly connected by a link, $l^{ij}\ni L$: $u^i(g+l^{ij}) < u^i(g)$ and $u^j(g+l^{ij}) < u^j(g)$
 \end{enumerate}
\end{Definition}
where the notation $g+l^{ij}$ denotes the network $g$ with the added link $l^{ij}$ and $g-l^{ij}$ the network with the link $l^{ij}$ removed. When maintaining a link is costly, there will be some network density that depends on the cost $c>0$ per link. The marginal utility of an additional link decreases with the number of links because the expected utility is bounded by $1$ (the pay-off is $1$ and the probability of choosing the correct action is less than, or equal to, $1$).\\

The algorithm to ensure a pairwise stable equilibrium starts by choosing a random agent $i$ from the set of agents $N$. Then, choose a second agent $j$ from the set of agents $N\setminus K^i$ that are not yet neighbors of $i$. Agents are chosen with the following probability:
\begin{equation}
w^j = \exp(\beta E^j)/Z,
\end{equation}
where $Z = \sum_k w^k$ is a normalization constant and $E^j = |\fez - \mu_0^j|$ is a proxy for agent $j$'s signal strength. For $\beta = 0$ $i$ chooses the new agent with equal probability. While this makes it more likely that agent $i$ considers forming a link with agent $j$ when $j$ has a higher signal strength, it does not imply that such a link is actually formed. This decision is solely based on the utility that both $i$ and $j$ obtain from establishing the link.\\

Now, let $K'^i = K^i\cup j$, i.e. the neigborhood of agent $i$ after adding adding agent $j$, and similarly $K'^j =  K^j \cup i$. The marginal utilities of adding $j$ and $i$ to the respective neighborhoods are then:
\begin{equation}
\begin{aligned}
	\Delta \bar{u}^i(q', K'^i, K^i) & = \bar{u}^i(q', K'^i) - \bar{u}^i(q', K^i) \\
	\Delta \bar{u}^j(q', K'^j, K^j) & = \bar{u}^j(q', K'^j) - \bar{u}^j(q', K^i) 
\end{aligned}
\end{equation}
Given the marginal utilities of agents $i$ and $j$ and their cost of maintaining link $c^i$ and $c^j$ the agents will form a link if $\Delta \bar{u}^i(q', K'^i, K^i) > c^i$ and $\Delta \bar{u}^j(q', K'^j, K^j) > c^j$. If $\Delta \bar{u}^i(q', K'^i, K^i) > c^i$ and $\Delta \bar{u}_j(q', K'^j, K^j) < c^j$, the algorithm selects the least informative agent in the neighborhood of $j$:
\begin{equation}
l = \underset{m \in K^j}{\operatorname{argmin}} E^m . 
\end{equation}
Now, define $K''^j = K'^j \setminus k$. If $\Delta \bar{u}^i(q', K'^i, K^i) > c^i$ and $\Delta \bar{u}^j(q', K''^j, K^j) > c^j$, form the link $l^{ij}$ and remove the link $l^{jk}$ (and similarly for $i\rightarrow j$ and $j\rightarrow i$). Otherwise, don't form the link. If $\Delta \bar{u}^i(q',K'^i,K^i) < c^i$ and $\Delta \bar{u}^j(q', K'^j, K^j) < c^j$ repeat the previous step, i.e. consider removing the least informative neighbor and re-evaluate the utilities.

\subsection{Equilibrium Networks}\label{Sec::EndogenousNetwork:Results}
The endogenously formed network in an economy with identically informed agents and positive cost $c$ of maintaining a link is a simple Erd\"os-R\'enyi network with a network density depending on the signal structure and link cost. To analyze more realistic situations, we can harvest the strengths of multi-agent simulations. In the following we therefore assume that agents are heterogenously informed about the underlying state of the world: a few ``informed'' agents have relatively precise signals and low costs of maintaining a link, while many ``uninformed'' agents have relatively imprecise signals and higher cost of maintaining a link. Table \ref{Table::OverviewParamtersEndo} summarizes the parameters we are using for the rest of this section.\\

In order to compare the dynamics on the endogenous networks to the ER networks we run the following simulations. We first create $1,000$ networks using the network formation algorithm described above. Then, we run the social learning algorithm described in Section \ref{Sec::ExogenousNetwork} while holding the network structure constant throughout. The underlying assumption is that banks are updating their investment decisions faster than the network structure changes. This can be empirically corroborated by looking at the term structure of interbank lending. While $90\%$ of the turnover in interbank markets is overnight, about $90\%$ of exposures between banks stems from the term segments.\\

We also run the social learning algorithm with an initialization bias in which we set the initial action of all agents to some pre-defined value. To assess the efficiency of the endogenous network formation, we compare the performance of the endogenously formed networks to the performance of Erd\"os-R\'enyi networks. All simulations in this section are conducted using the equal weighting scenario. An example of a resulting network structure can be found in Figure \ref{FIG::EXAMPLE_GRAPH_1} and the degree distribution of the endogenously formed networks in $1,000$ is shown in Figure \ref{FIG::ENDO_DEGREE_DIST}. The degree distribution of the endogenously formed networks is clearly bimodal. One peak corresponds to the uninformed nodes with small degree while the second peak corresponds to the informed nodes with high degree. Figure \ref{FIG::ENDO_X_FINAL_DIST} shows the distribution of the final action for the $1,000$ simulations conducted. A clear improvement over the Erd\"os-R\'enyi networks can be seen, highlighting the importance of core banks with more precise private signals. Since core banks are highly interconnected, there is a higher chance that they are in the neighborhood of a peripheral bank (as opposed to a peripheral bank being in the neighborhood of another peripheral bank) which increases the precision of peripheral banks' social belief.\\

To further understand the difference between endogenously formed and random networks, we analyze the time it takes learning to converge. We assume the learning has converged at time $t$ if:
\begin{equation}
	\sum_i| x_i(t) - x_i(t+\Delta t) | / N < 0.05,\textrm{ where } \Delta t = 15.
\end{equation}
It can be seen from Figure (\ref{FIG::ENDO_CONV_T_DIST}) that, except for very long convergence times, the system always converges faster in the endogenous network case than in the Erd\"os-R\'enyi case. Note, that this simulation was conducted without initialization bias, i.e. with average initial action of $\fez$. Finally, the probability of contagion as a function of an initialization bias, i.e. as a function of average initial action is shown in Figure (\ref{FIG::ENDO_P_CONT_LOG}). Again, the picture is unanimously showing that the probability of contagion, i.e. the probability that more than $80\%$ of agents coordinate on a state non-matching action is significantly smaller in the endogenous network case than in the case of a random Erd\"os-R\'enyi graph.

\section{Conclusion}\label{Sec::Conclusion}
This paper develops a model of contagious synchronization of bank's investment strategies. Banks are connected via mutual lines of credit and endogenously choose an optimal network structure. They receive a private signal about the state of the world and observe the strategies of their counterparties. When banks observe the actions of more peers they put more weight on their social belief. We compare three scenarios of weighting functions. First, in the equal weighting scenario, agents place equal weights on their private and social belief. Second, in the neighborhood scenario agents place proportionately more weight on the social signal when the size of the neighborhood increases. Third, in the relative neighborhood scenario agents place more weight on the social belief if their neighborhood constitutes a larger fraction of the overall network. Social learning increases the probability of choosing a state matching action and thus agents' utility. When agents strategically choose their neighbors they take the additional utility from learning into account. The more neighbors a given agent has, the lower is the marginal utility from another link and the network endogenously reaches an equilibrium configuration.\\

We obtain two results which are policy relevant. First, in a complex financial system where agents cannot take the action of all their peers into account when taking an investment decision, the probability of contagious synchronization depends on two things: (i) the weighting between the private and social belief; and (ii) the density of the financial network. Our model thus relates two empirically relevant sources of systemic risk: common shocks interbank market freezes. Second, the probability of contagious synchronization is substantially reduced when agents internalize the positive effects of social learning in a strategic decision with whom to form a link. The benefit from learning is reduced when the private signals about the state of the world are less informative.\\

The model has a number of interesting extensions. One example is the case with two different regions that can feature differing states of the world. Such an application could capture a situation in which banks in two countries (one in a boom, the other in a bust) can engage in interbank lending within the country and across borders. This would provide an interesting model for the current situation within the Eurozone. The model so far features social learning but not individual learning. Another possible extension would be to introduce individual learning and characterize the conditions under which the contagious regime exists. Finally, the model can be applied to real-world interbank network and balance sheet data to test for the interplay of contagious synchronization and endogenous network structure.\\

One drawback of the model is that there is no closed-form analytical solution for the benefit a bank obtains through learning from a peer that takes into account higher order effects. This benefit will depend on whether or not a neighboring bank chose a state matching on state non-macthing action in the previous period and thus on the social belief of neighboring banks. In the former case, the benefit will be positive, while in the latter case it will be negative. Agents have ex ante no way of knowing what action a neighboring bank selected until the state of the world is revealed ex post. Finding such a closed-form solution is beyond the scope of the present paper which focuses on the application in an agent-based model, but would provide a fruitful exercise for future research.

\clearpage
\bibliographystyle{model2-names}
\bibliography{bibliography_c}
\clearpage
\clearpage
\newpage
\pagenumbering{roman}
\setcounter{page}{1}
\begin{appendix}
\section{Tables}\label{Appendix::Tables}

\begin{table}[h]
\begin{center}
\small
\begin{tabular}{llrrr}
\toprule
Variable & Description & \multicolumn{3}{c}{Value} \\
\cmidrule{3-5}
&  & 1 - (I) & 2 - (U) & 3 - (H) \\
\midrule
$N$ & Number of agents & $100$ & $100$ & $100$ \\ 
$\mu_0$ & Average signal for $\theta = 0$ & $0.4$ & $0.49$ & $0.3$ \\ 
$\mu_1$ & Average signal for $\theta = 1$ & $0.6$ & $0.51$ & $0.7$ \\ 
$\sigma_0$ & Standard deviation of signal for $\theta = 0$ & $\sqrt{0.1}$ & $\sqrt{0.1}$ & $\sqrt{0.1}$ \\ 
$\sigma_1$ & Standard deviation of signal for $\theta = 1$ & $\sqrt{0.1}$ & $\sqrt{0.1}$ & $\sqrt{0.1}$ \\ 
$T$ & Number of iterations of updater & $100$ & $100$ & $100$ \\ 
$\rho$ & Density of ER network & $[0,0.95]$ & $[0,0.95]$ & $0.5$ \\ 
$p$ & Probability of being informed & NA & NA & $[0.1,0.9]$ \\ 
$S$ & \# of simulations per param. config. & $1000$ & $1000$ & $1000$ \\ 
\bottomrule 
\end{tabular}
\end{center}
\caption{Parameters for runs for the case of informed (I), uninformed (U), and heterogeneous (H) agents.}\label{Table::OverviewParamters}
\end{table}

\begin{table}[h]
\begin{tabular}{llr} 
\toprule
Notation & Description & Value \\
\midrule
$N$ & Number of agents & $30$\\
$N_I$ & Number of informed agents & $4$\\
$N_U$ & Number of unformed agents & $26$\\
$\mu_{0I}$ & Average signal for $\theta = 0$ for informed agents& $0.3$\\
$\mu_{1I}$ & Average signal for $\theta = 1$ for informed agents& $0.7$\\
$\sigma_{0I}$ & Standard deviation of signal for $\theta = 0$ for informed agents& $\sqrt{0.1}$ \\
$\sigma_{1I}$ & Standard deviation of signal for $\theta = 1$ for informed agents& $\sqrt{0.1}$ \\
$\mu_{0U}$ & Average signal for $\theta = 0$ for uninformed agents& $0.4$\\
$\mu_{1U}$ & Average signal for $\theta = 1$ for uninformed agents& $0.7$\\
$\sigma_{0U}$ & Standard deviation of signal for $\theta = 0$ for uninformed agents& $\sqrt{0.1}$ \\
$\sigma_{1U}$ & Standard deviation of signal for $\theta = 1$ for uninformed agents& $\sqrt{0.1}$ \\
$c_{I}$ & Cost per link for informed agent& $0$\\
$c_{U}$ & Cost per link for informed agent& $0.1$\\
$T_C$ & Number of iterations of network algorithm & $400$ \\
$q'$ & Agent's belief of average action of neighbors of neighbors & $0.5$ \\
$\beta$ & Intensity of choice in agent selection & $30$ \\
\midrule
$T$ & Number of iterations of action updater & $100$ \\ 
$n$ & Number of endogenously formed networks used for simulation & $1000$ \\ 
$S$ & Number of simulations per parameter configuration & $100$ \\
$\rho$ & Average density of ER networks & $0.08$ \\ 
\bottomrule 
\end{tabular}
\caption{Parameters for network formation and runs with endogenous networks.}\label{Table::OverviewParamtersEndo}
\end{table}

\clearpage 

\section{Figures}\label{Appendix::Figures}
\begin{figure}[ht]
\centering
\begin{minipage}{0.5\textwidth}
	\includegraphics[width=\textwidth]{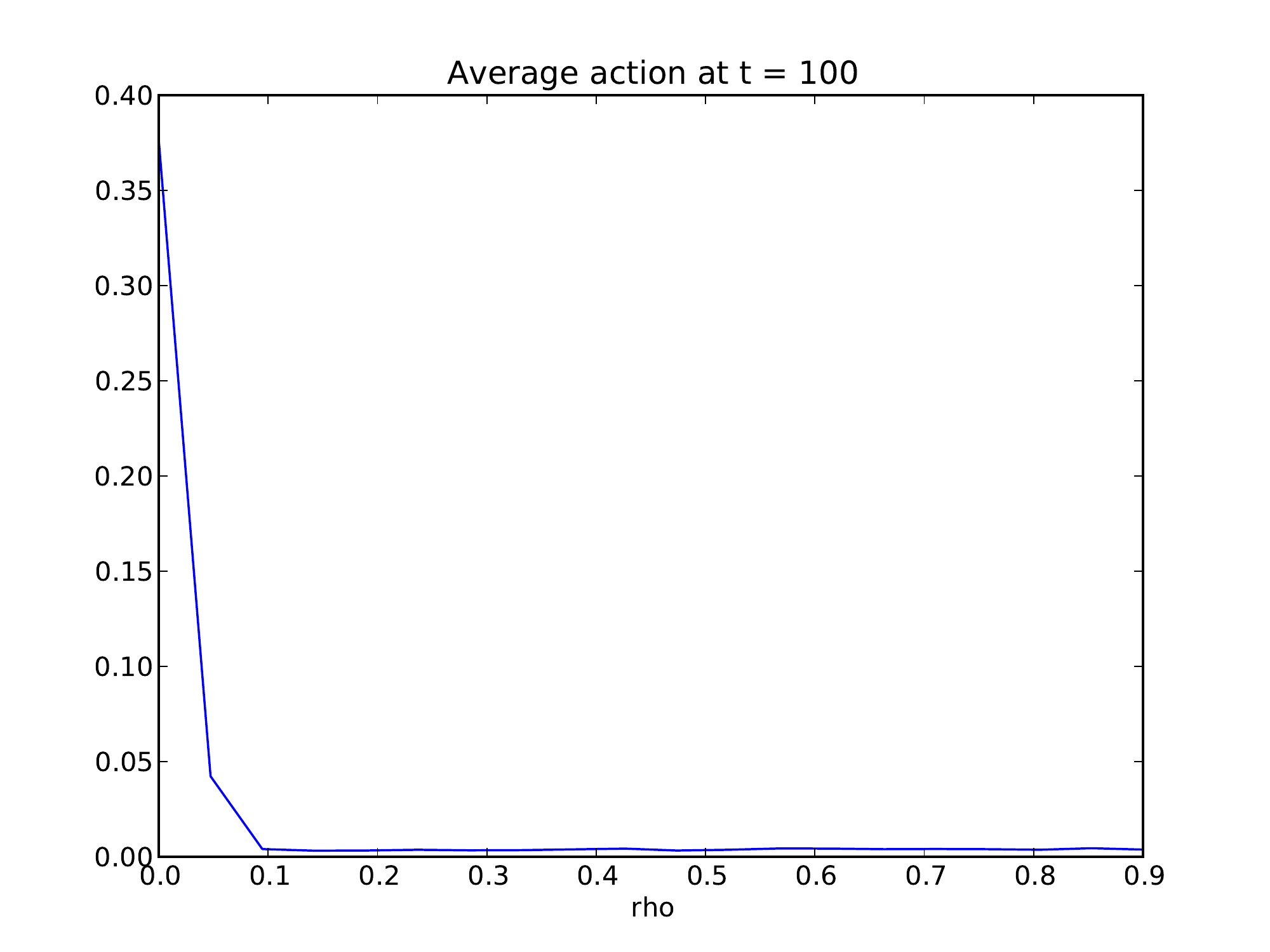}\\
	\includegraphics[width=\textwidth]{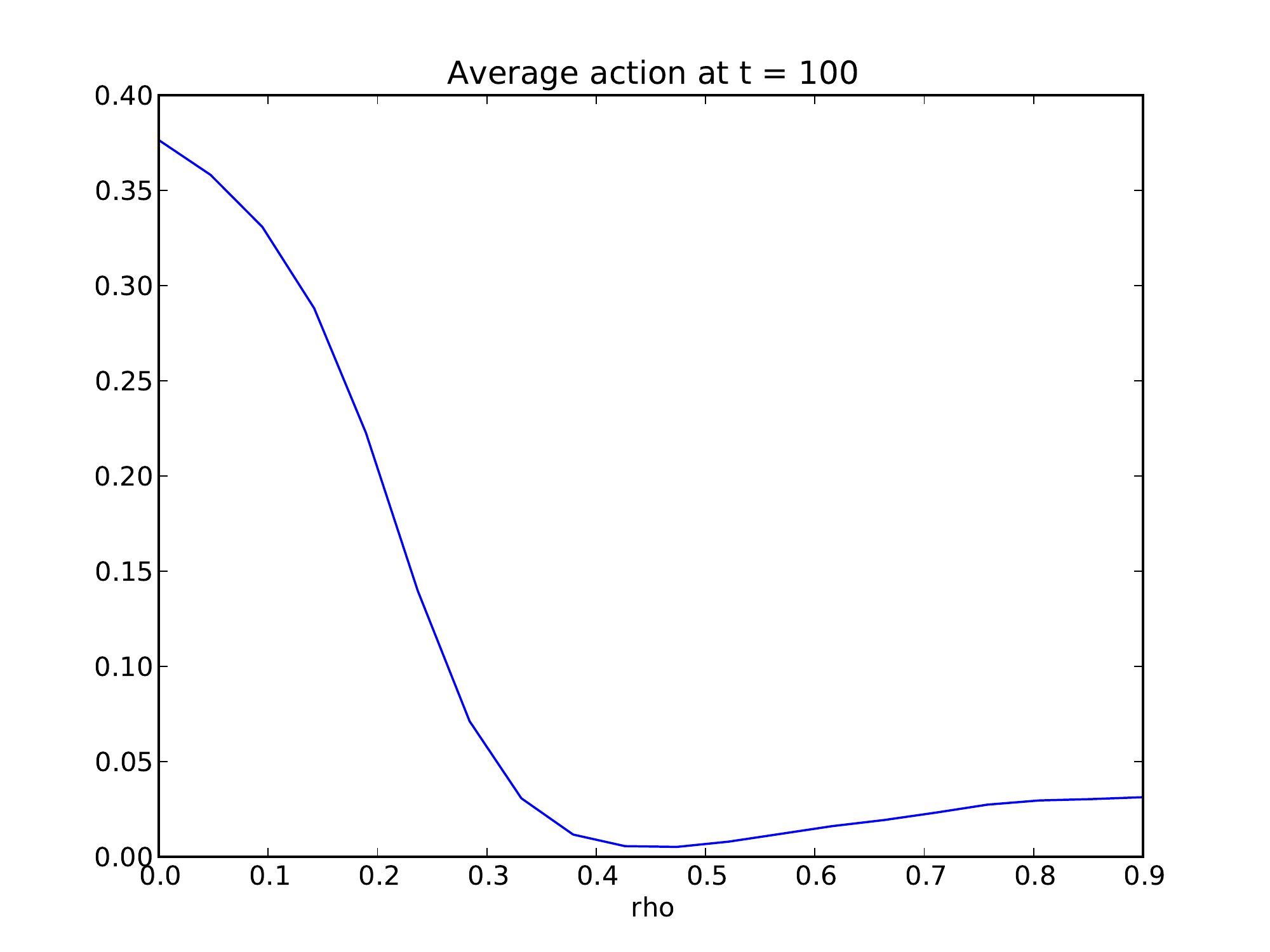}\\
	\includegraphics[width=\textwidth]{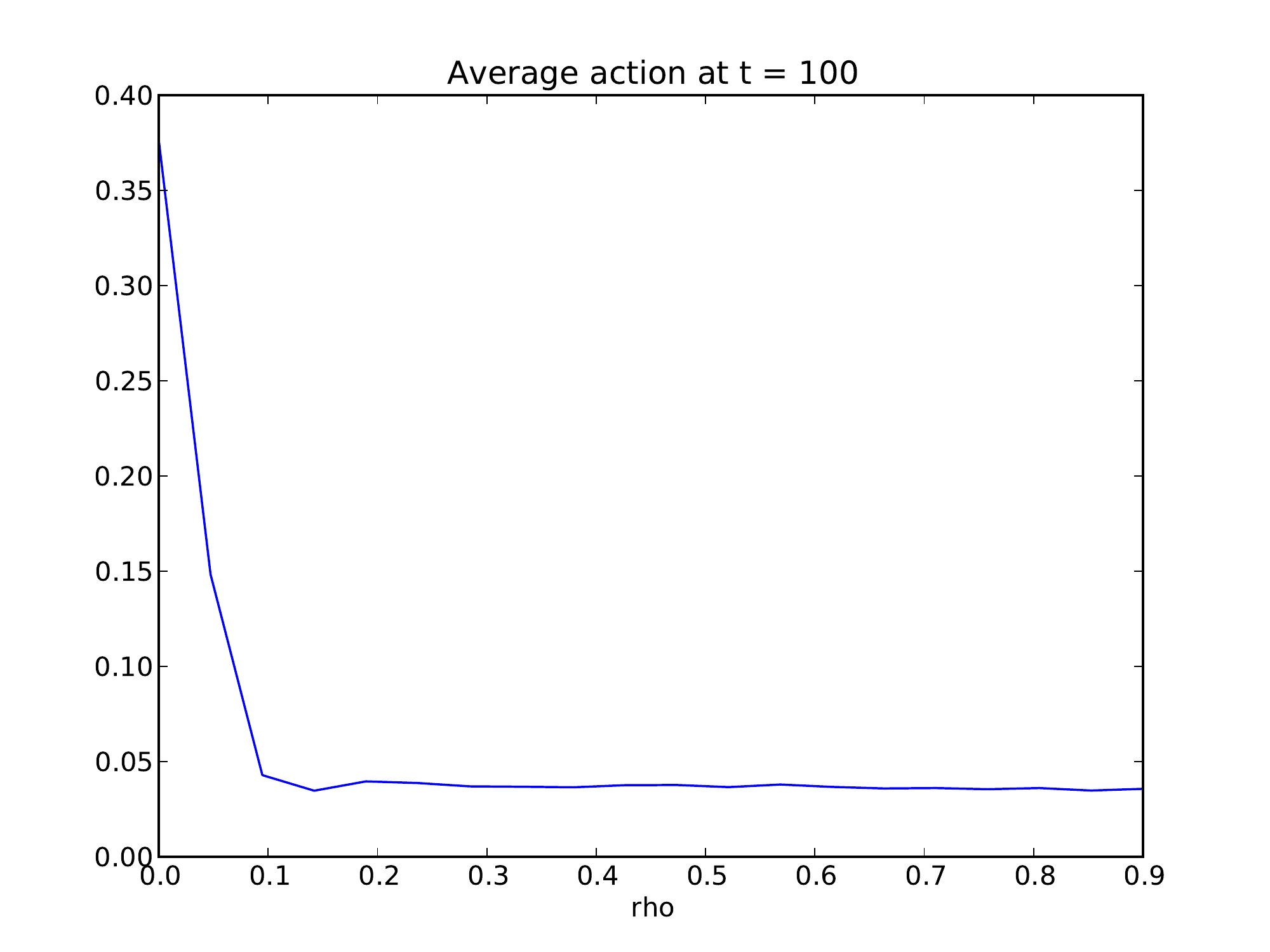}
\end{minipage}\begin{minipage}{0.5\textwidth}
	\includegraphics[width=\textwidth]{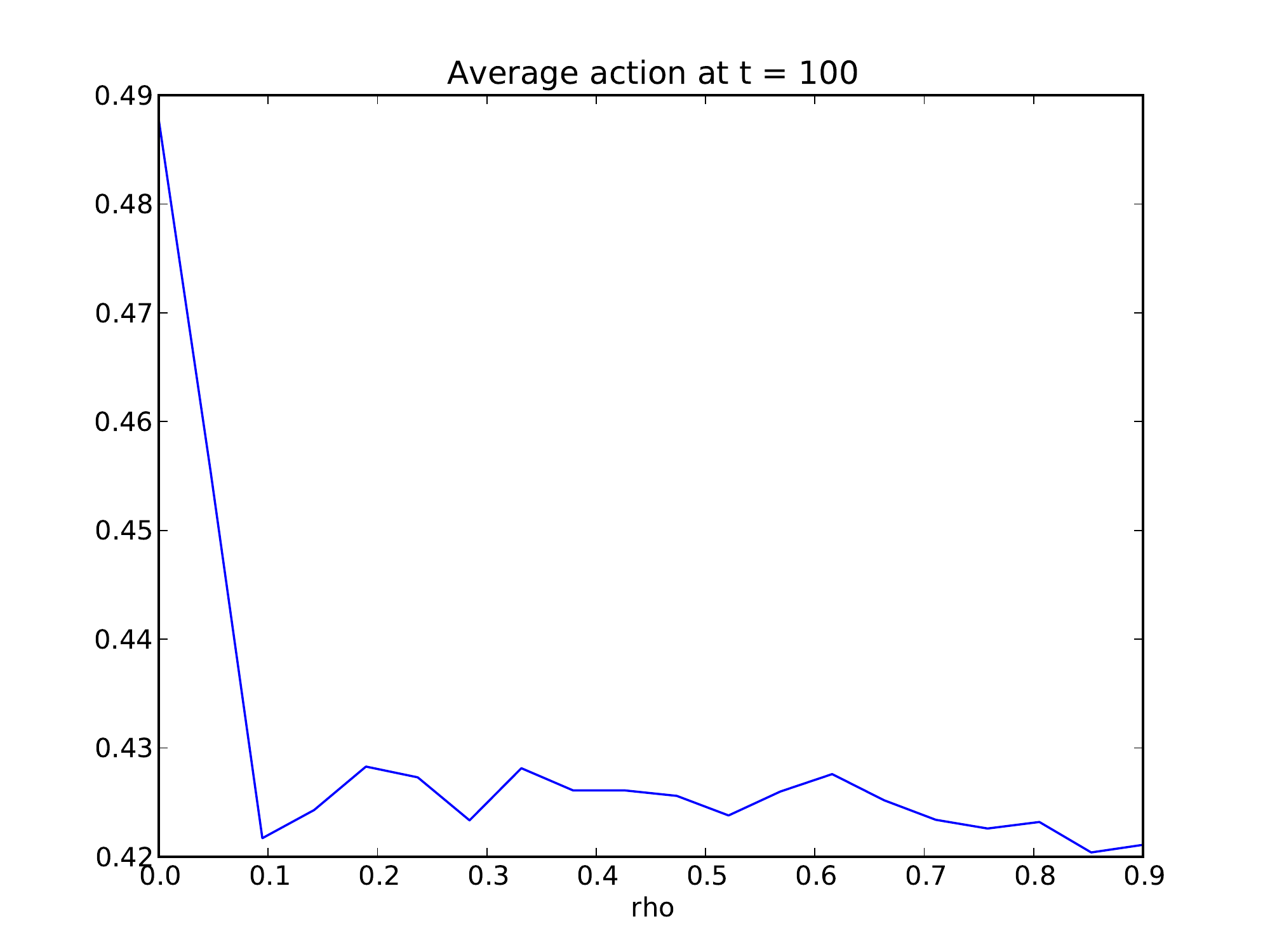}\\
	\includegraphics[width=\textwidth]{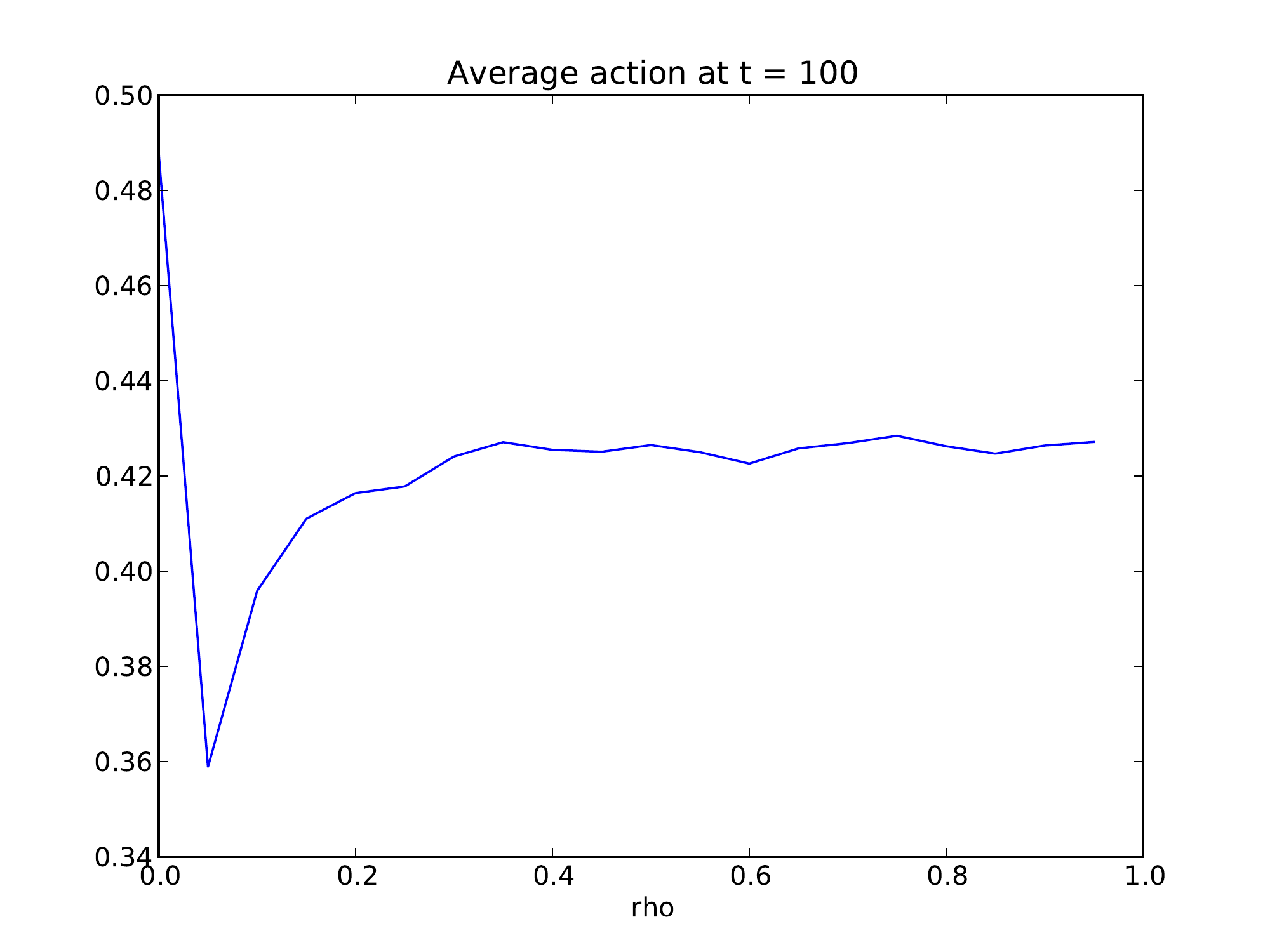}\\
	\includegraphics[width=\textwidth]{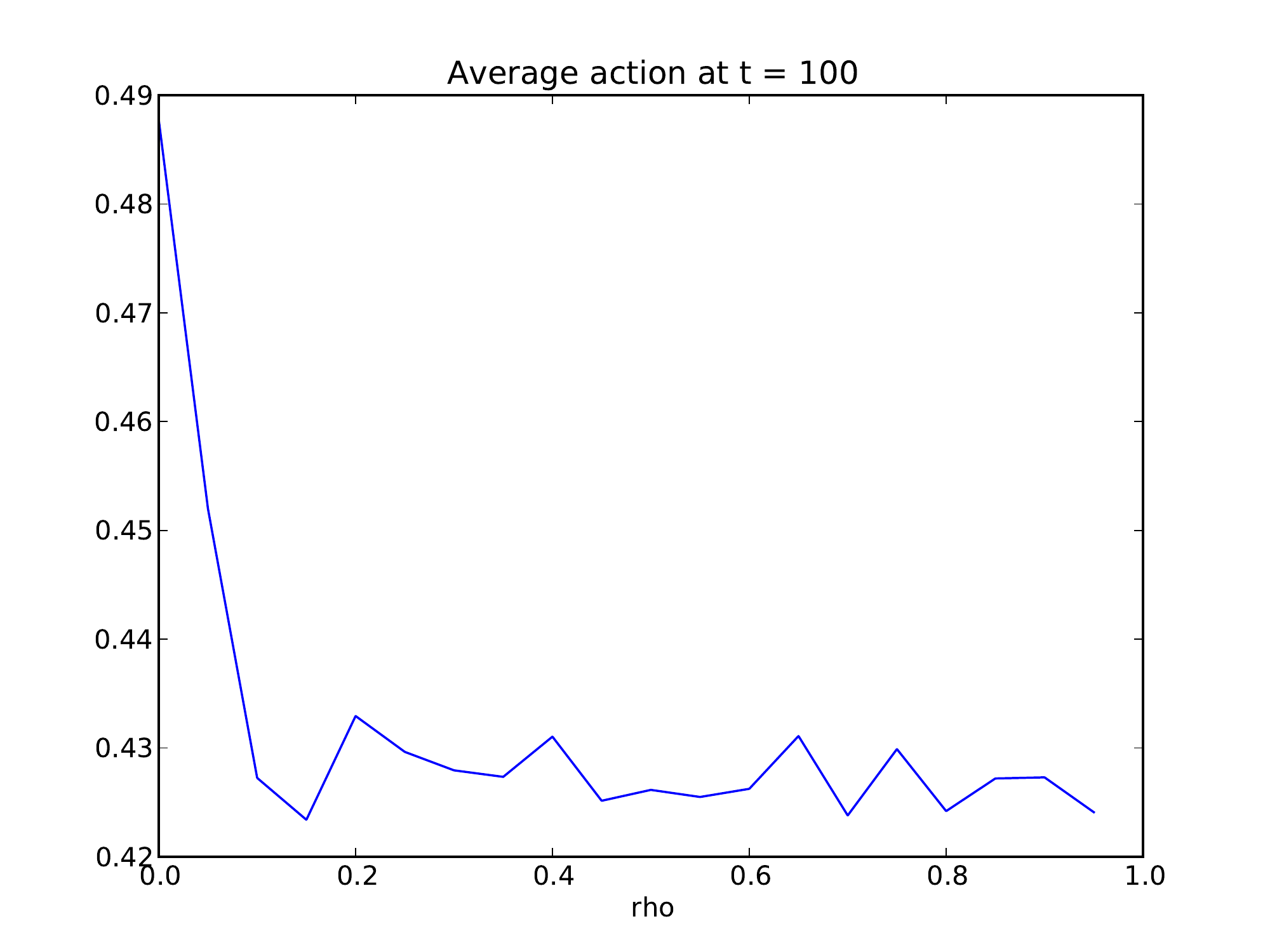}
\end{minipage}
\caption{Average final action of agents as a function of the network density rho of a random graph for the informed (left) and uninformed (right) case. Top: Equal weighting scenario; Center: Neighborhood size scenario; Bottom: relative neighborhood scenario.}\label{Figure::ExoAvgFinalAction}
\end{figure}

\begin{figure}[ht]
\centering
\begin{minipage}{0.5\textwidth}
	\includegraphics[width=\textwidth]{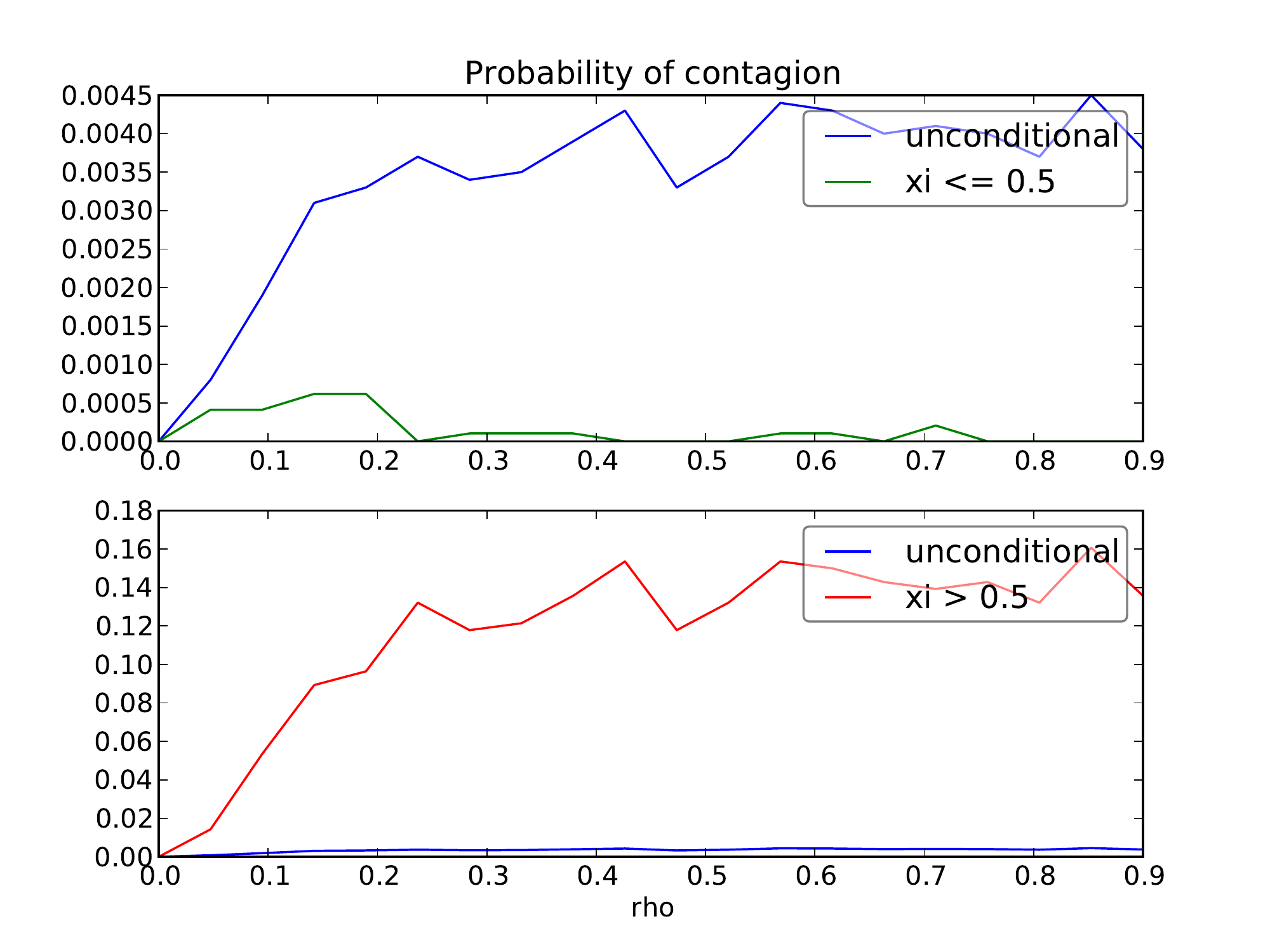}\\
	\includegraphics[width=\textwidth]{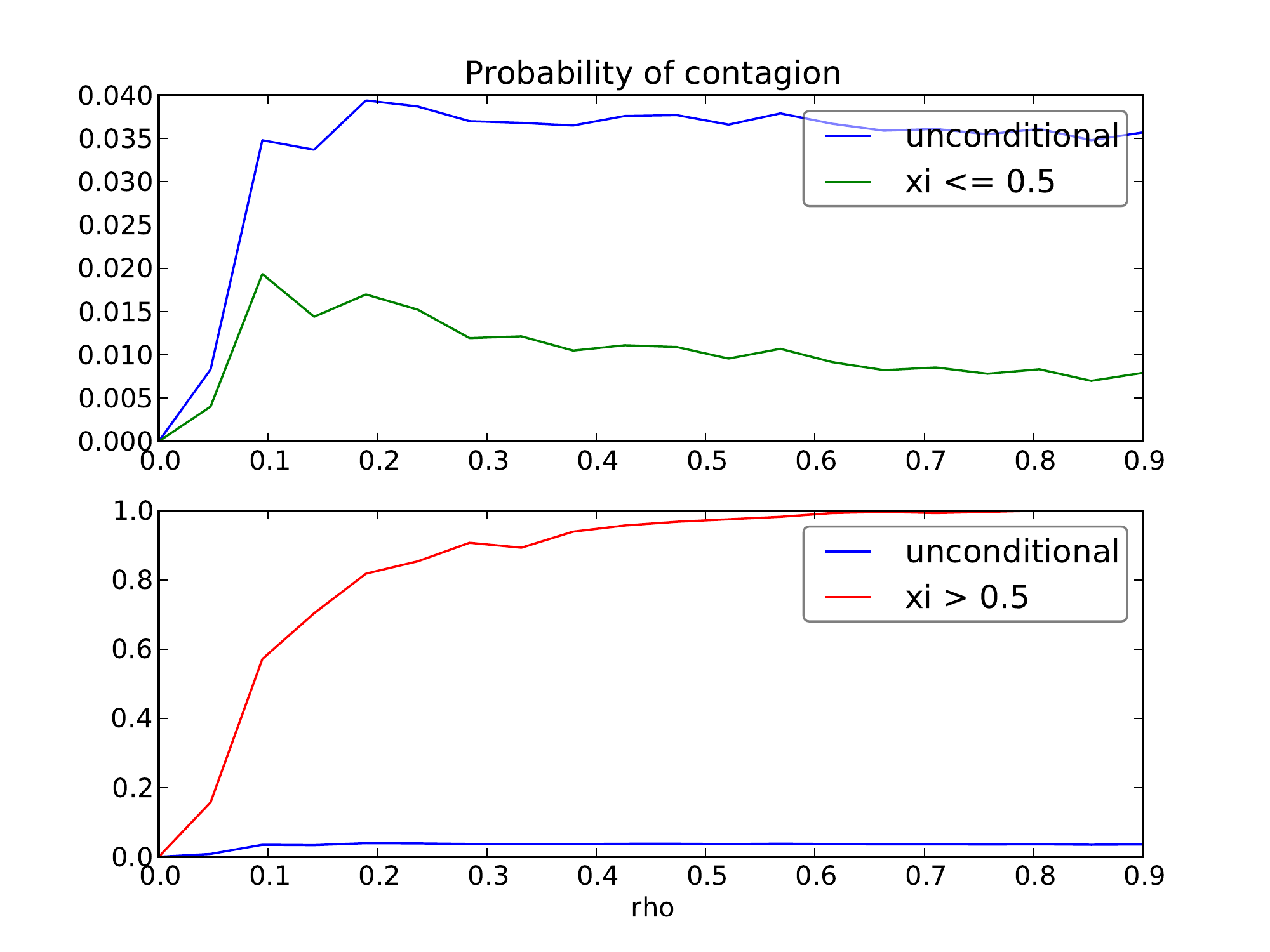}\\
	\includegraphics[width=\textwidth]{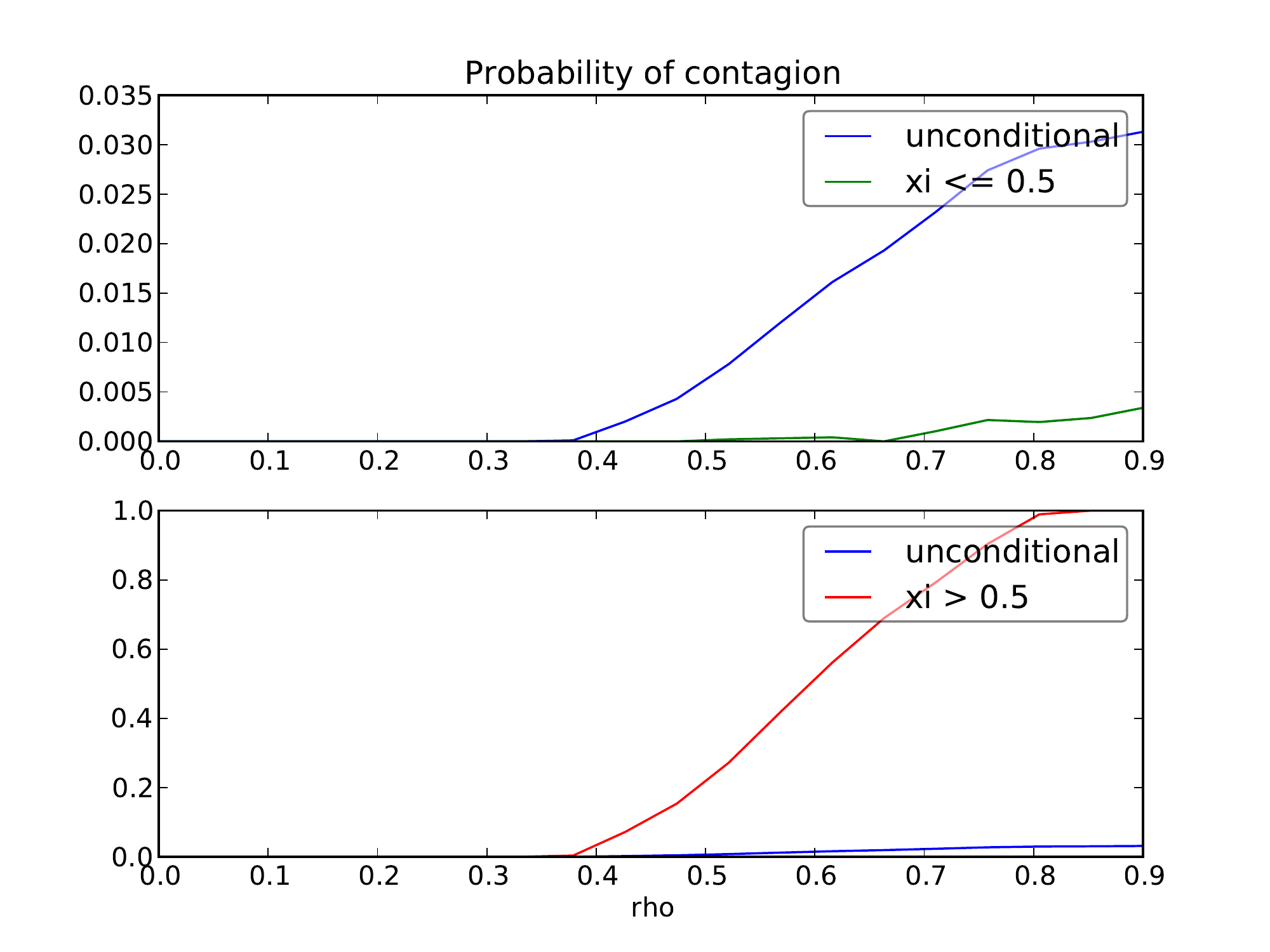}
\end{minipage}\begin{minipage}{0.5\textwidth}
	\includegraphics[width=\textwidth]{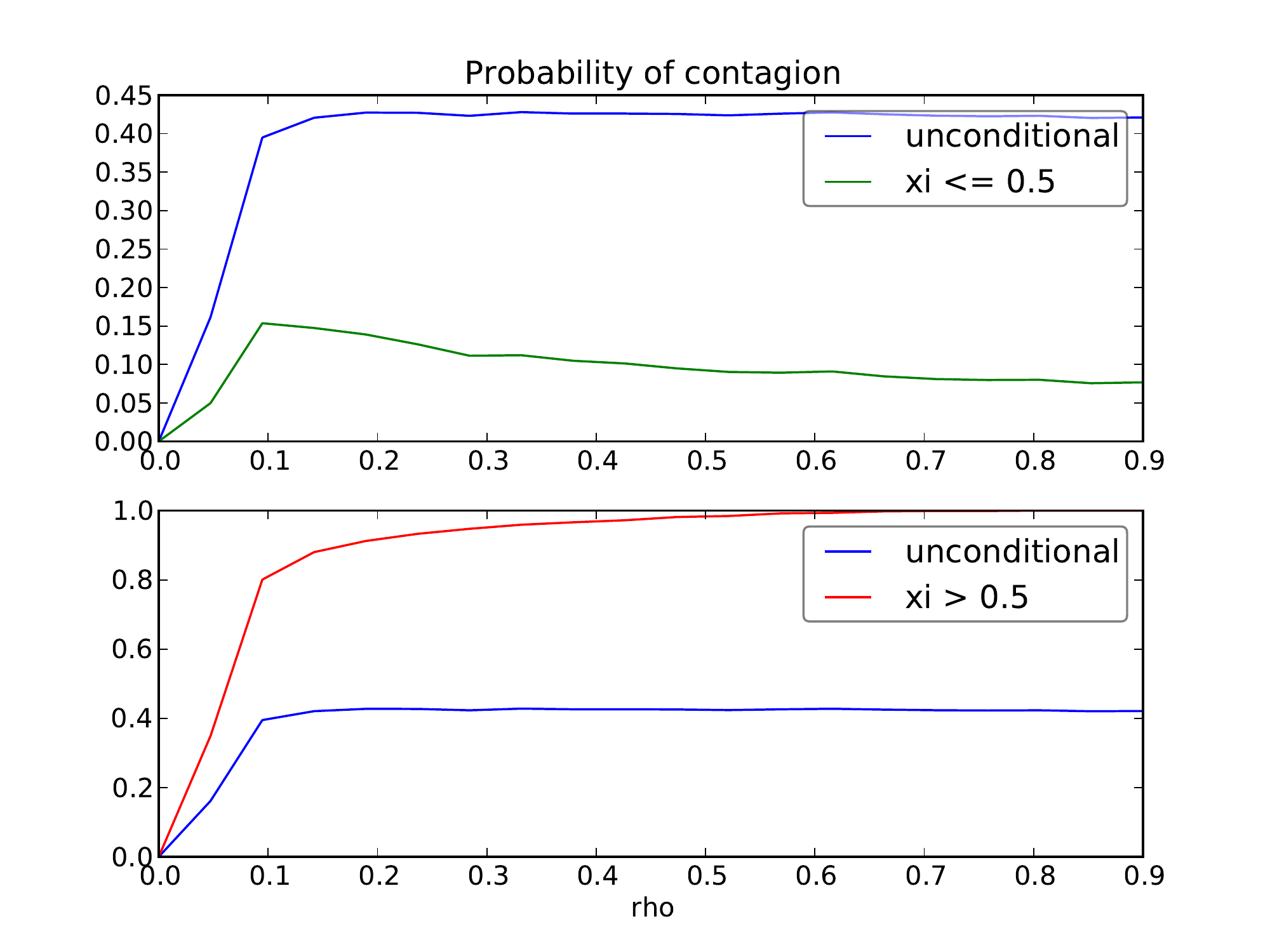}\\
	\includegraphics[width=\textwidth]{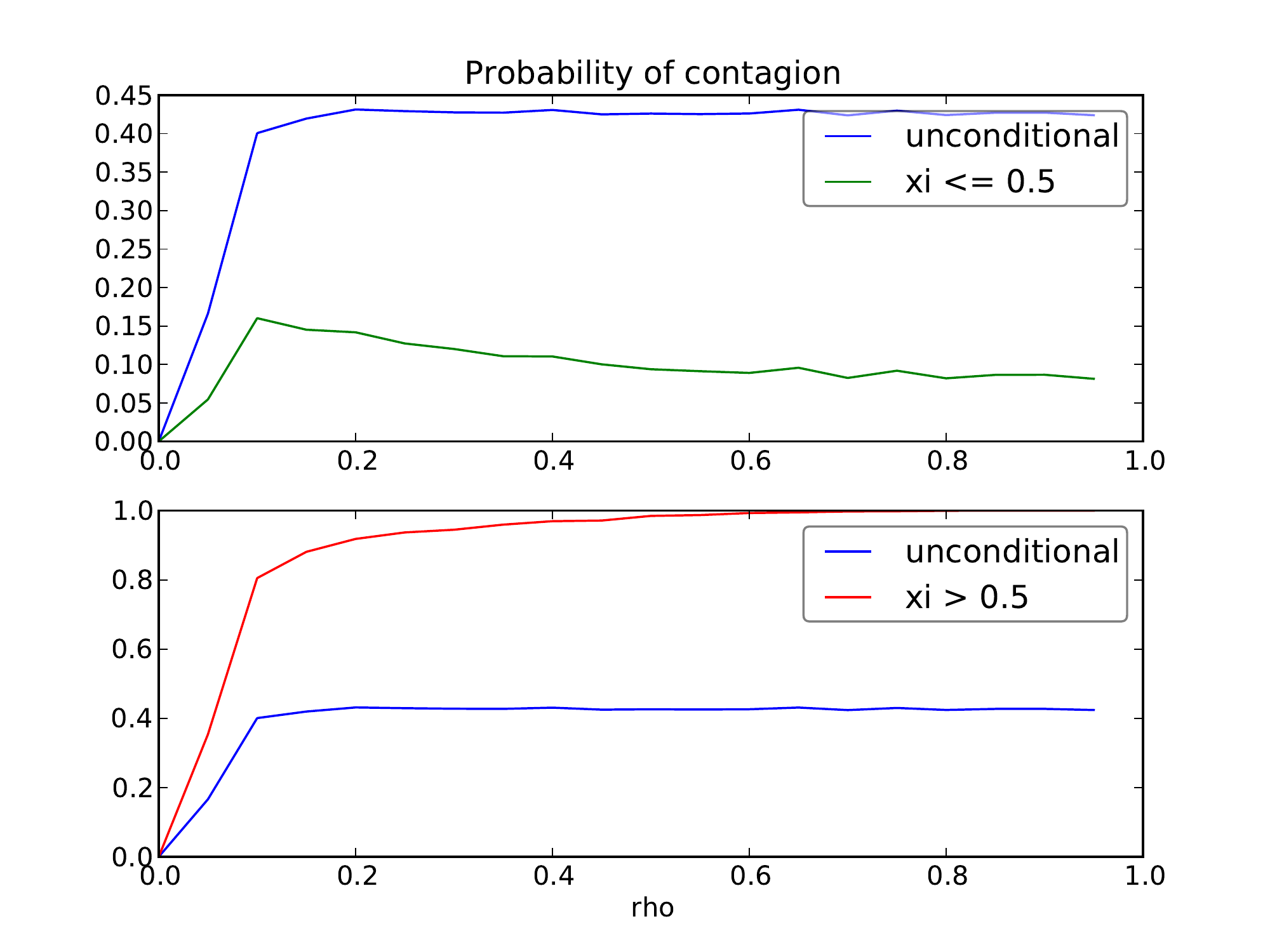}\\
	\includegraphics[width=\textwidth]{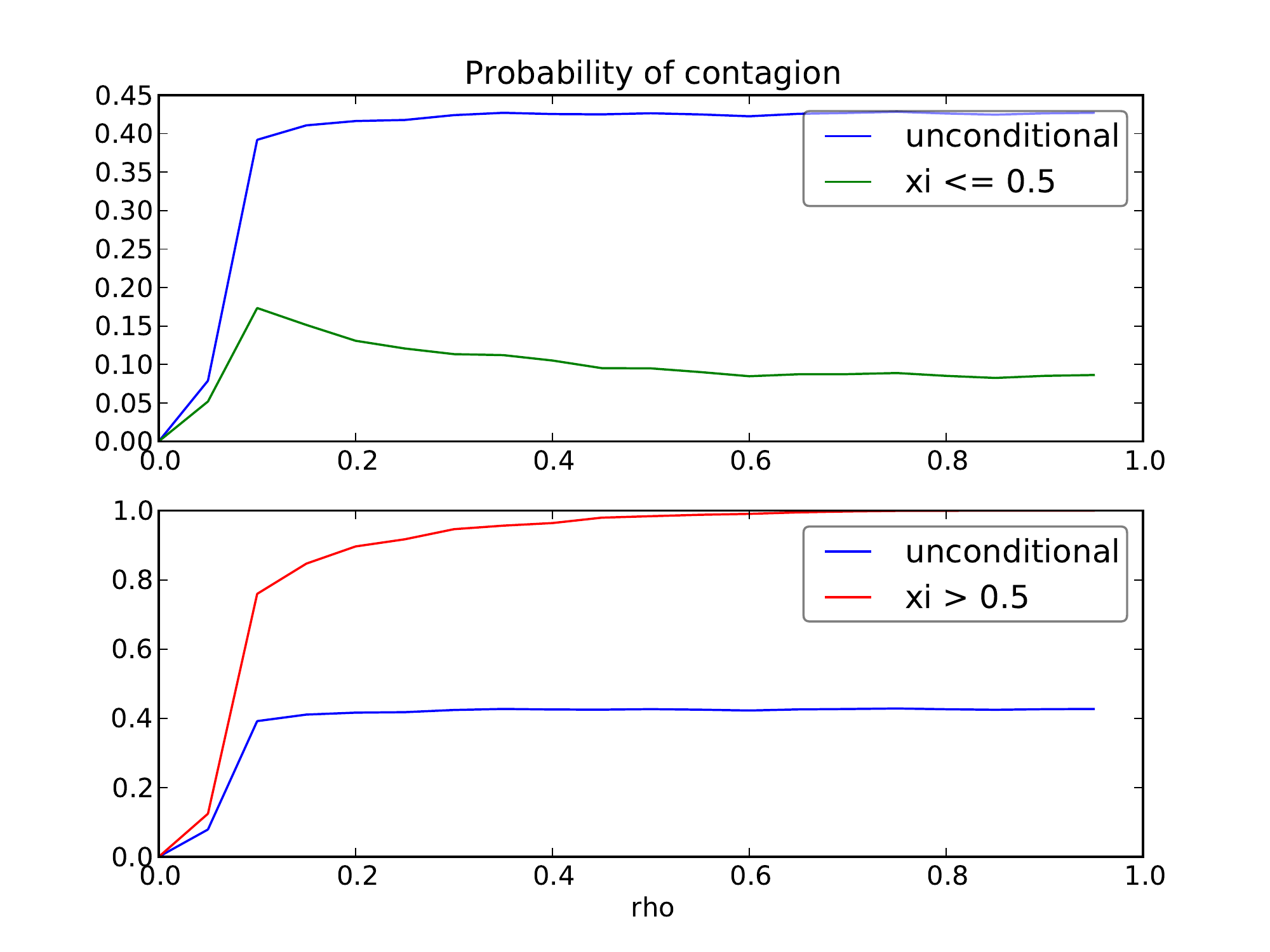}
\end{minipage}
\caption{Fraction of simulations per parameter configuration $S$ (1000) in which agents synchronize on the state non matching action (more than 80\% of agents choose state non-matching action) as a function of network density rho of a random graph for the informed (left) and uninformed (right) case. Top: Equal weighting scenario; Center: Neighborhood size scenario; Bottom: relative neighborhood scenario. We distinguish three cases: (1) unconditional: we compute the fraction based on the full sample $S$. (2) conditional $\hat{x} \leq \fez$: we compute the fraction based on the sub-set of simulations in which the average initial action $\hat{x} = \sum_i x^i(0)/N \leq \fez$, i.e. when the agents start with a state matching action. (3) conditional $\hat{x} > \fez$: we compute the fraction based on the sub-set of simulations when the agents start with a state non matching action.}\label{Figure::ExoFracFinalAction}
\end{figure}

\begin{figure}[ht]
\centering
\begin{minipage}{0.5\textwidth}
	\includegraphics[width=1.1\textwidth]{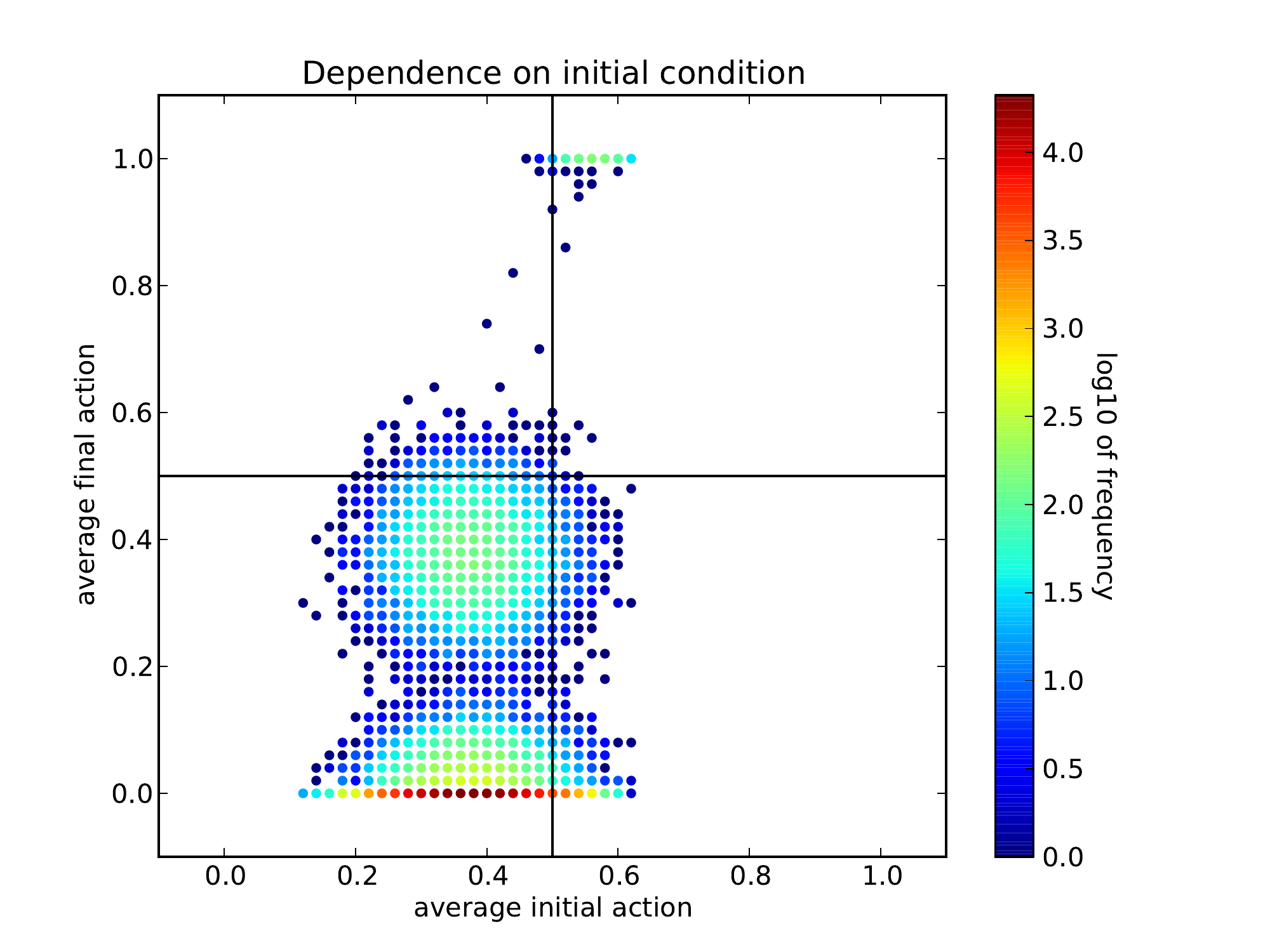}\\
	\includegraphics[width=1.1\textwidth]{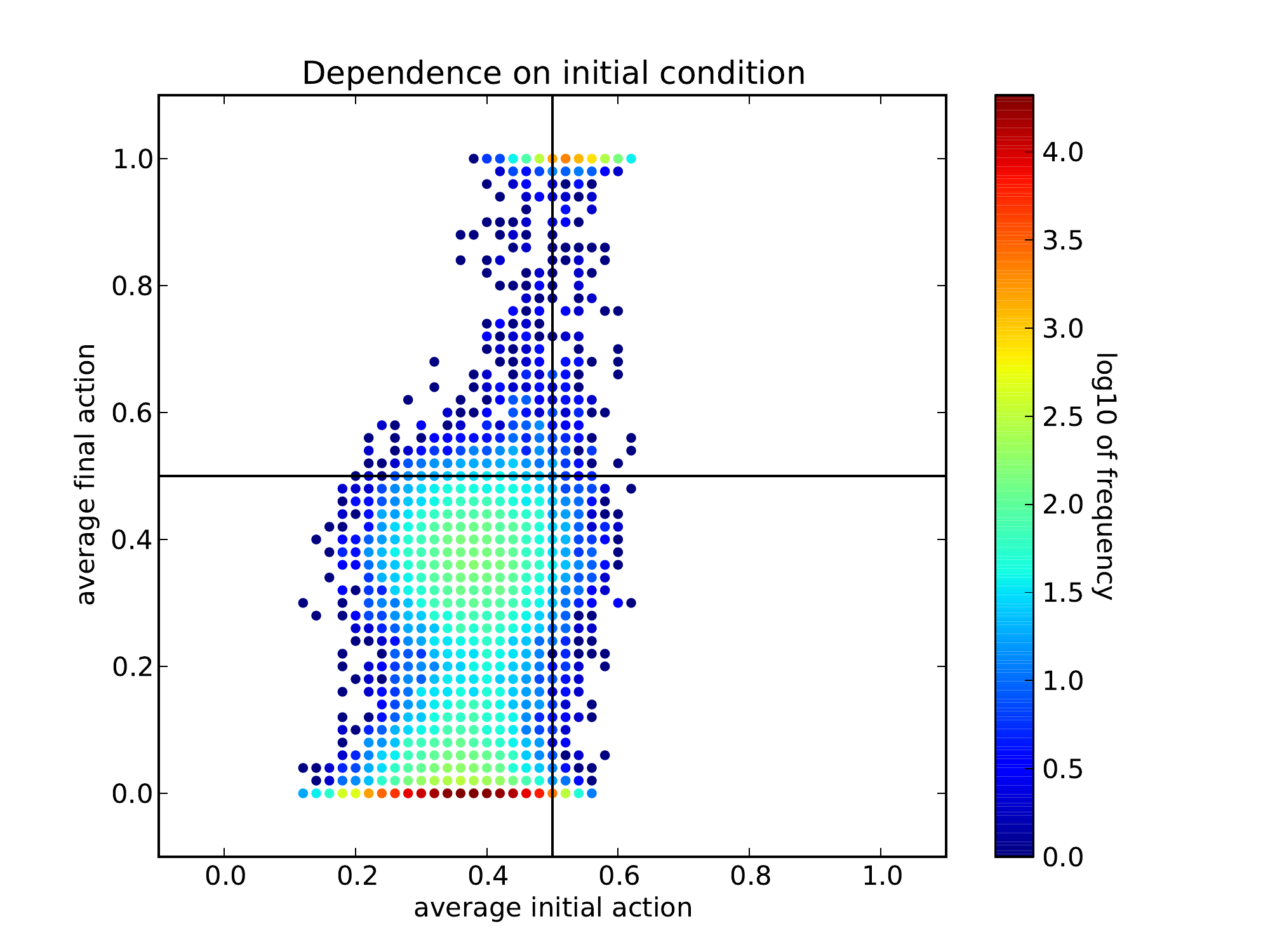}
	\includegraphics[width=1.1\textwidth]{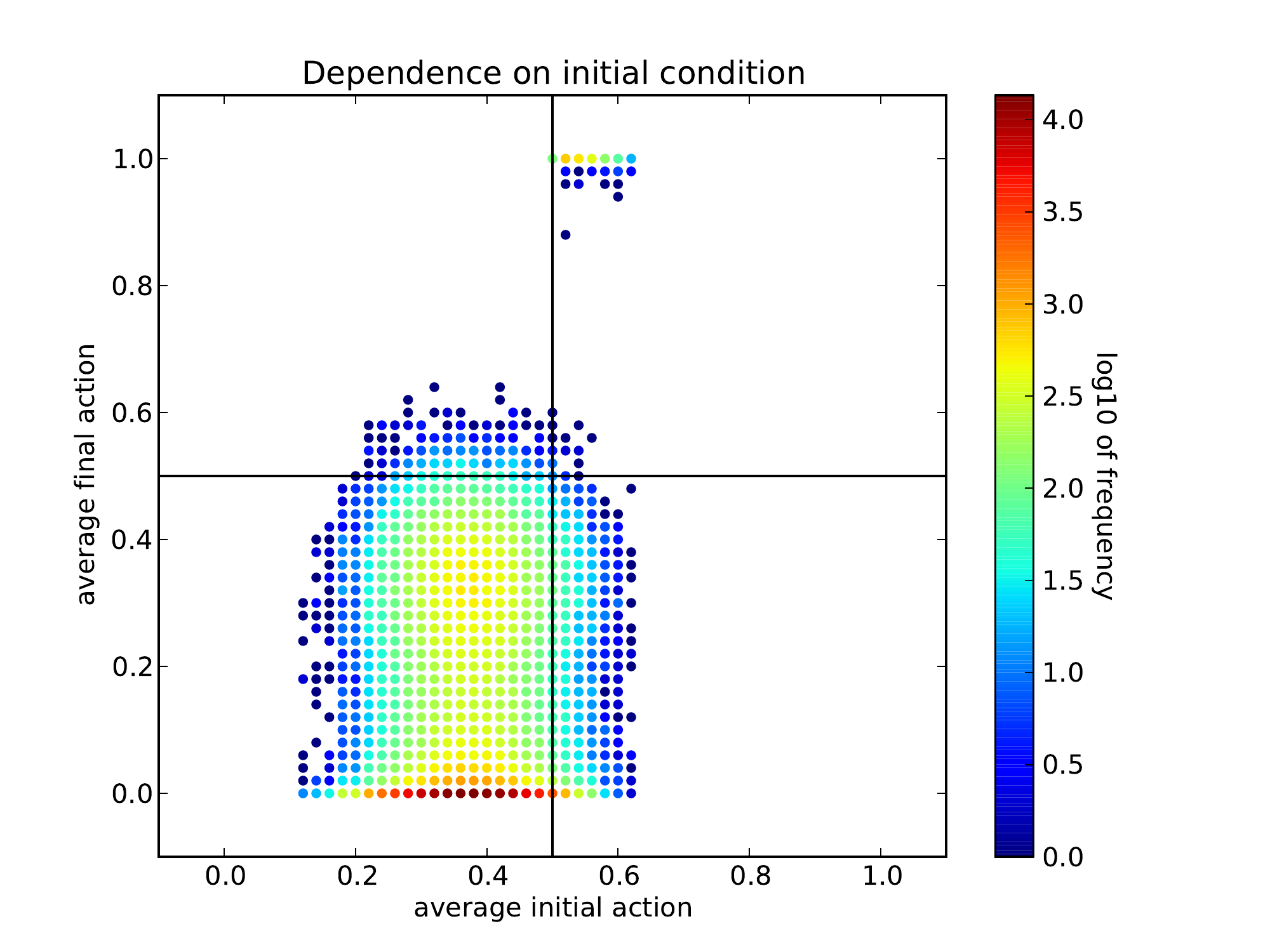}
\end{minipage}\begin{minipage}{0.5\textwidth}
	\includegraphics[width=1.1\textwidth]{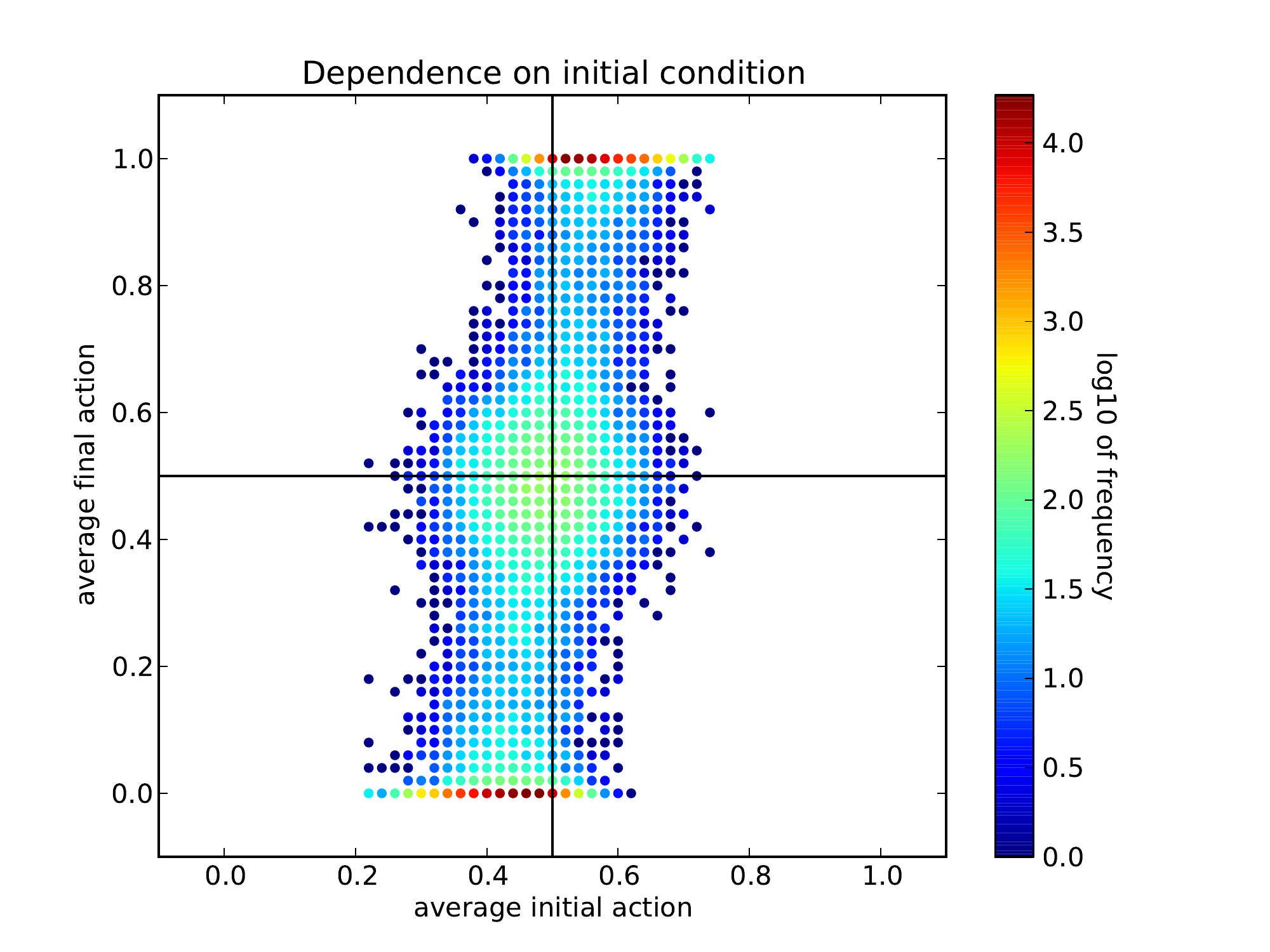}\\
	\includegraphics[width=1.1\textwidth]{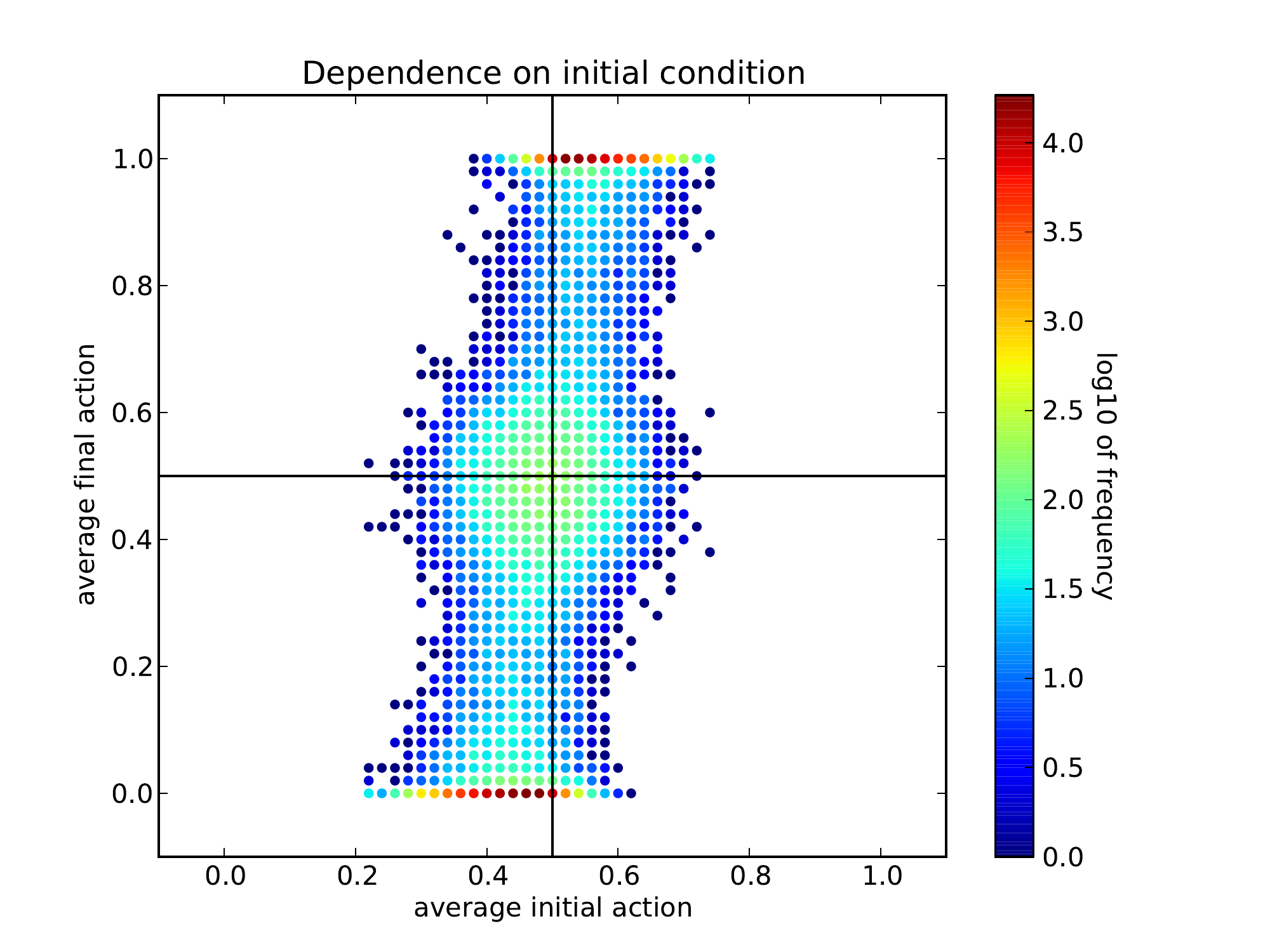}
	\includegraphics[width=1.1\textwidth]{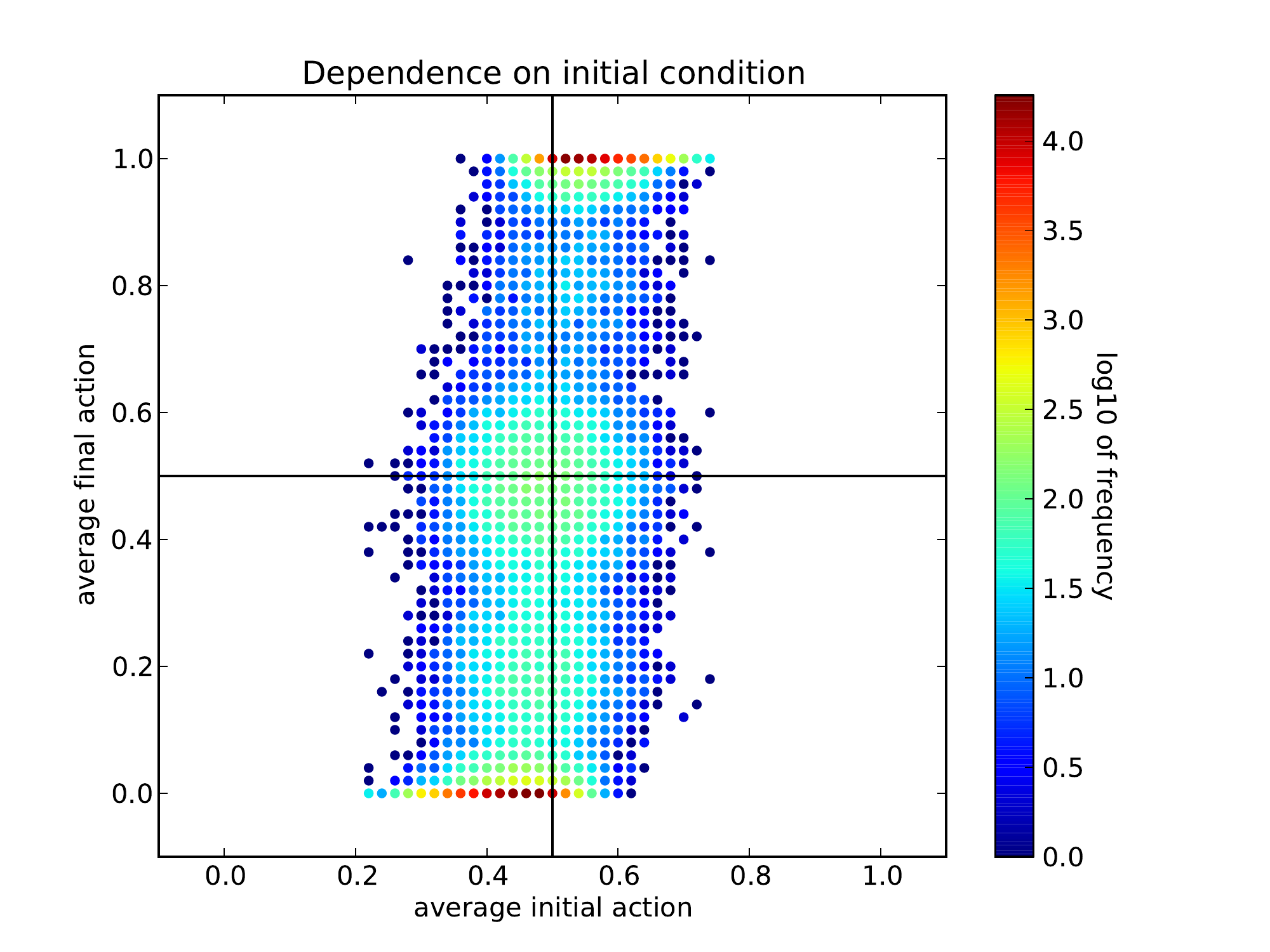}
\end{minipage}
\caption{Average final action $\hat{x}_F = \sum_i x^i(T)/N$ versus the average initial action $\hat{x}_I = \sum_i x^i(0)/N$ for the informed (left) and uninformed (right) case. Top: Equal weighting scenario; Center: Neighborhood size scenario; Bottom: relative neighborhood scenario. Data points are averages over $S = 1000$ simulations and all network densities $\rho$ (20 values equally distributed over the interval $[0,0.95]$). The color code indicates the frequency with which a point occurs in the sample (total size $20 \times 1000$), the scale of the color code is logarithmic of base 10.}\label{Figure::ExoAvgFinalVSAvgInitial}
\end{figure}

\begin{figure}[ht]
\centering
\includegraphics[width=0.7\textwidth]{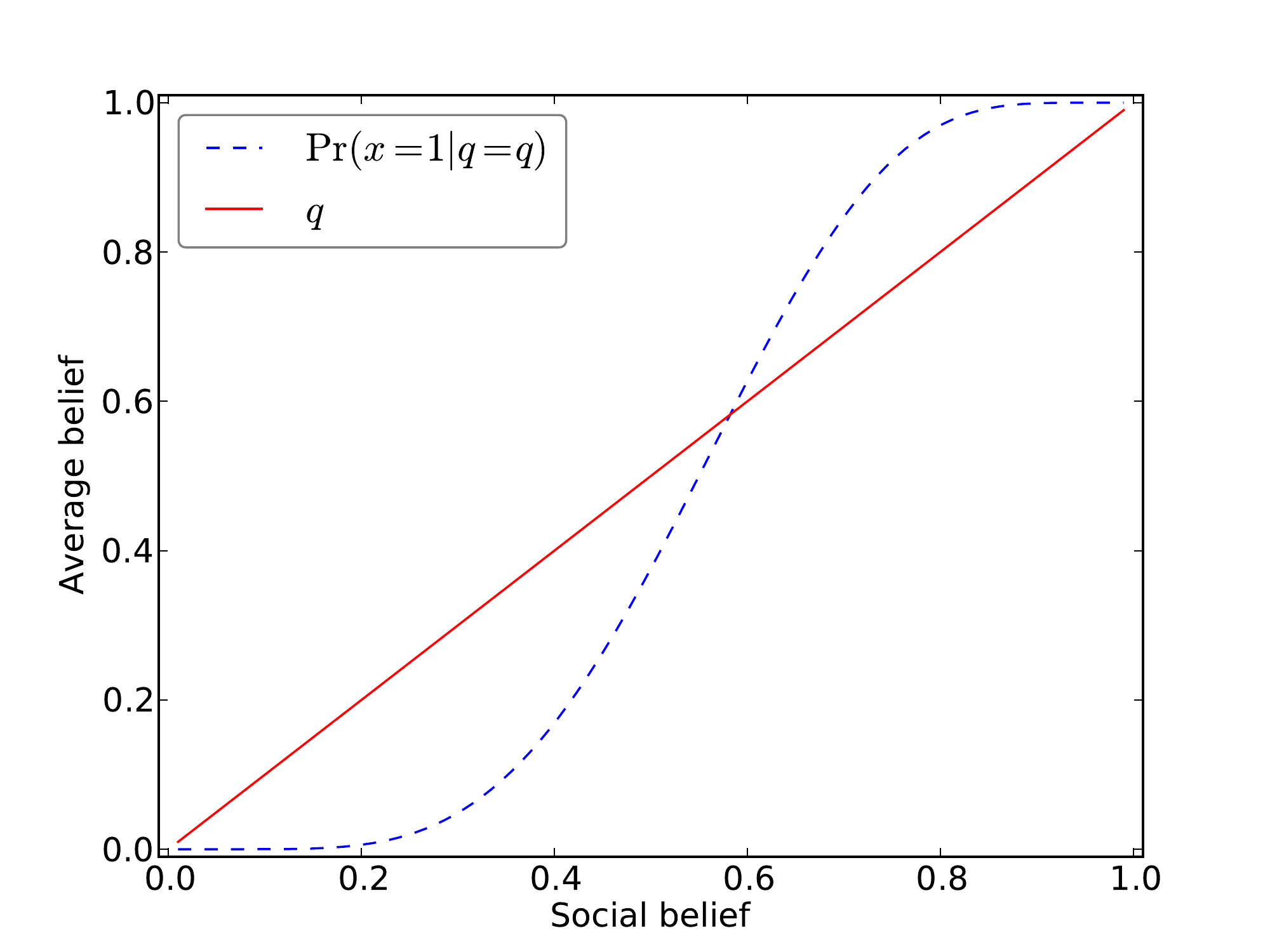}
\caption{Probability of choosing state matching action (equivalent to average social belief and average action in mean field) given that the state of the world is $\theta=0$, $f_p(p^i \mid \theta = 0)$ for $\mu_0 = 0.4$. Furthermore we use $\mu_1 = 1 - \mu_0$ and $\sigma^2 = 0.1$. The intersections between the probability function and the diagonal mark the fixed points of the dynamical system.}\label{Fig::EquilibriumSocialBelief}
\end{figure}

\begin{figure}[ht]
\centering
\includegraphics[width=\textwidth]{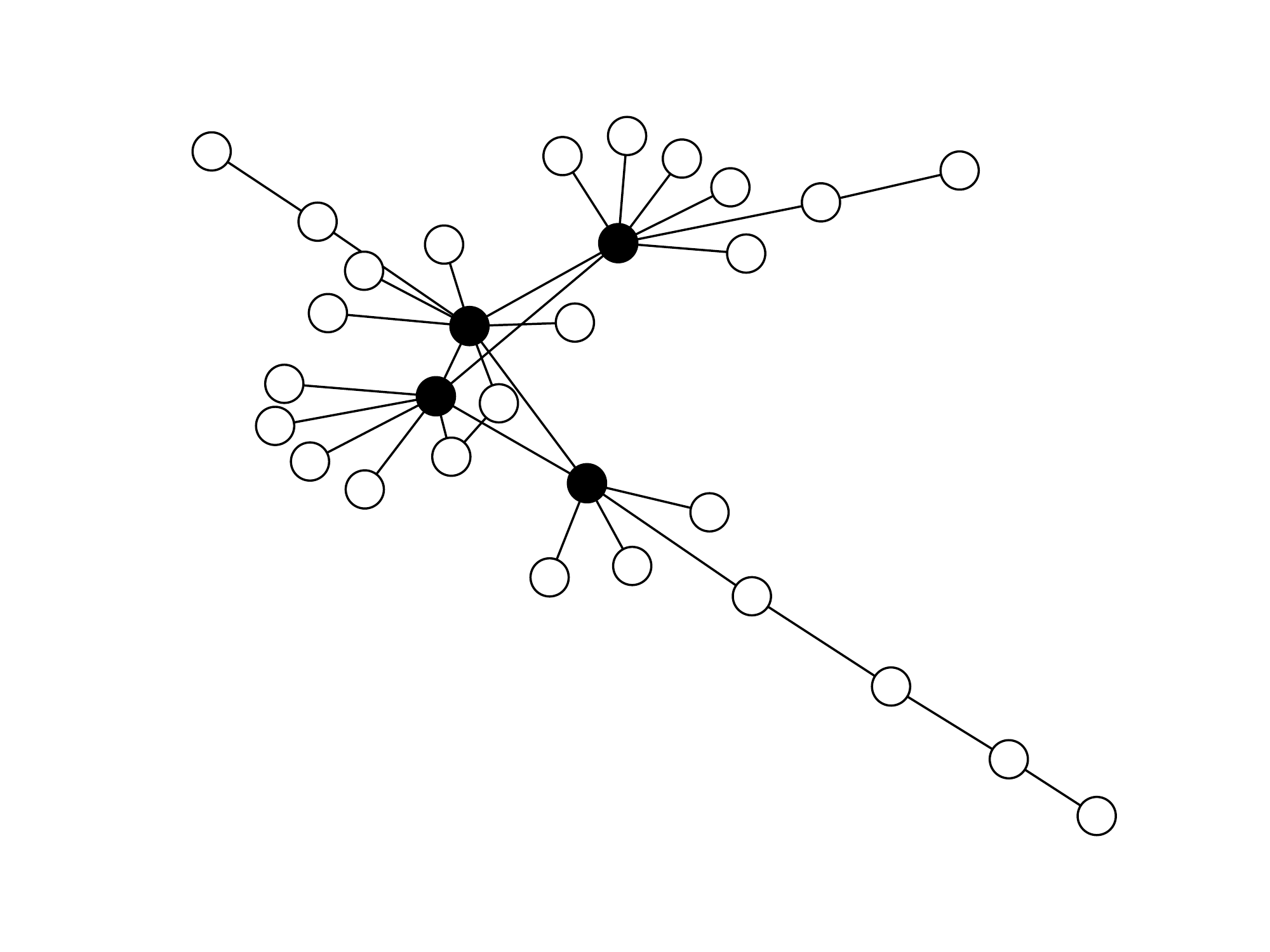}
\caption{Example network from ensemble of $n=1000$ endogenously formed networks. Black nodes are ``informed'' agents, while white nodes are ``uninformed''.}\label{FIG::EXAMPLE_GRAPH_1}
\end{figure}

\begin{figure}[ht]
\centering
\includegraphics[width=0.7\textwidth]{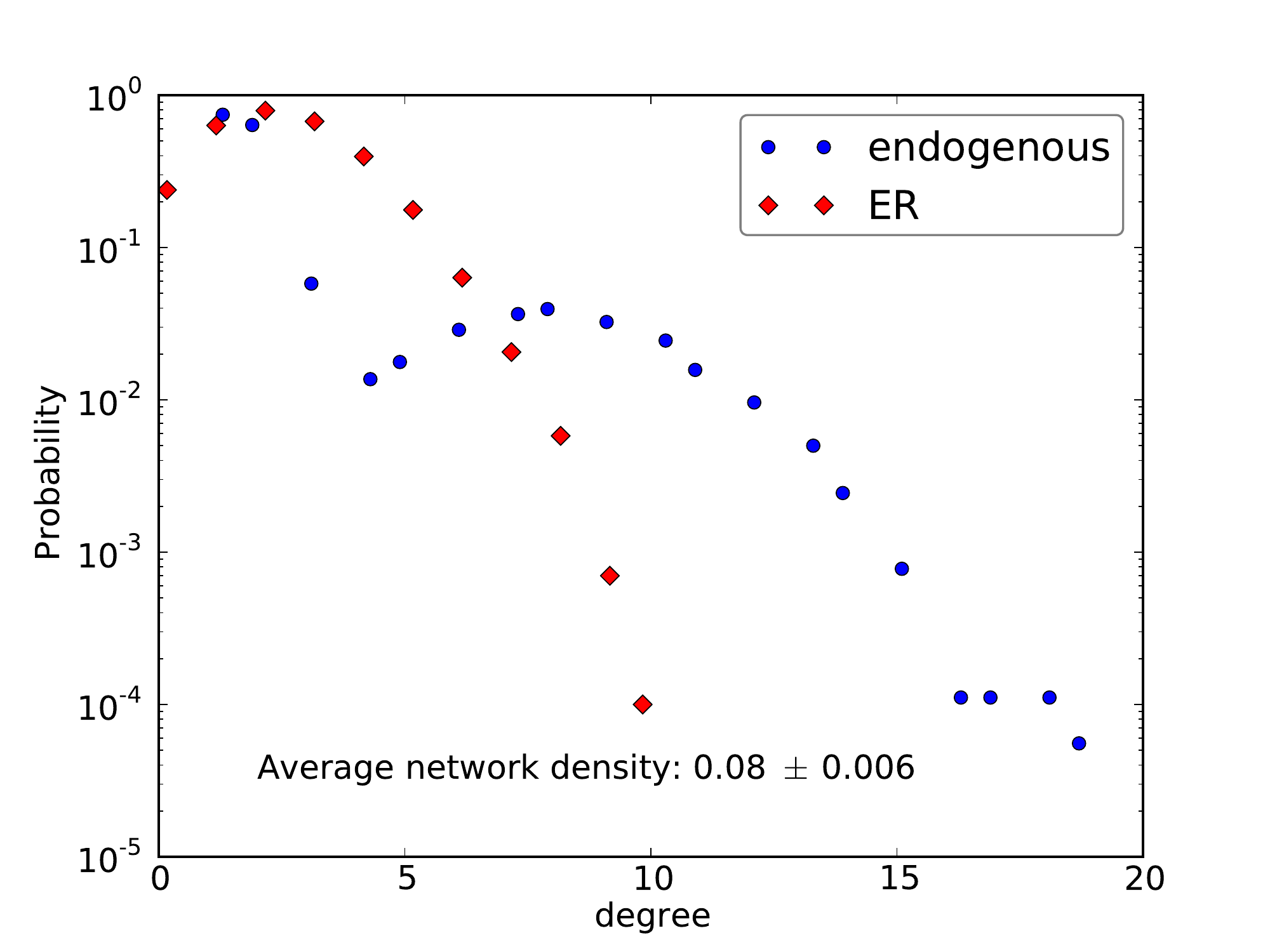}
\caption{Degree distribution of ensemble of $n=1000$ endogenously formed networks compared to ER networks with same average density.}\label{FIG::ENDO_DEGREE_DIST}
\end{figure}

\begin{figure}[ht]
\centering
\includegraphics[width=0.7\textwidth]{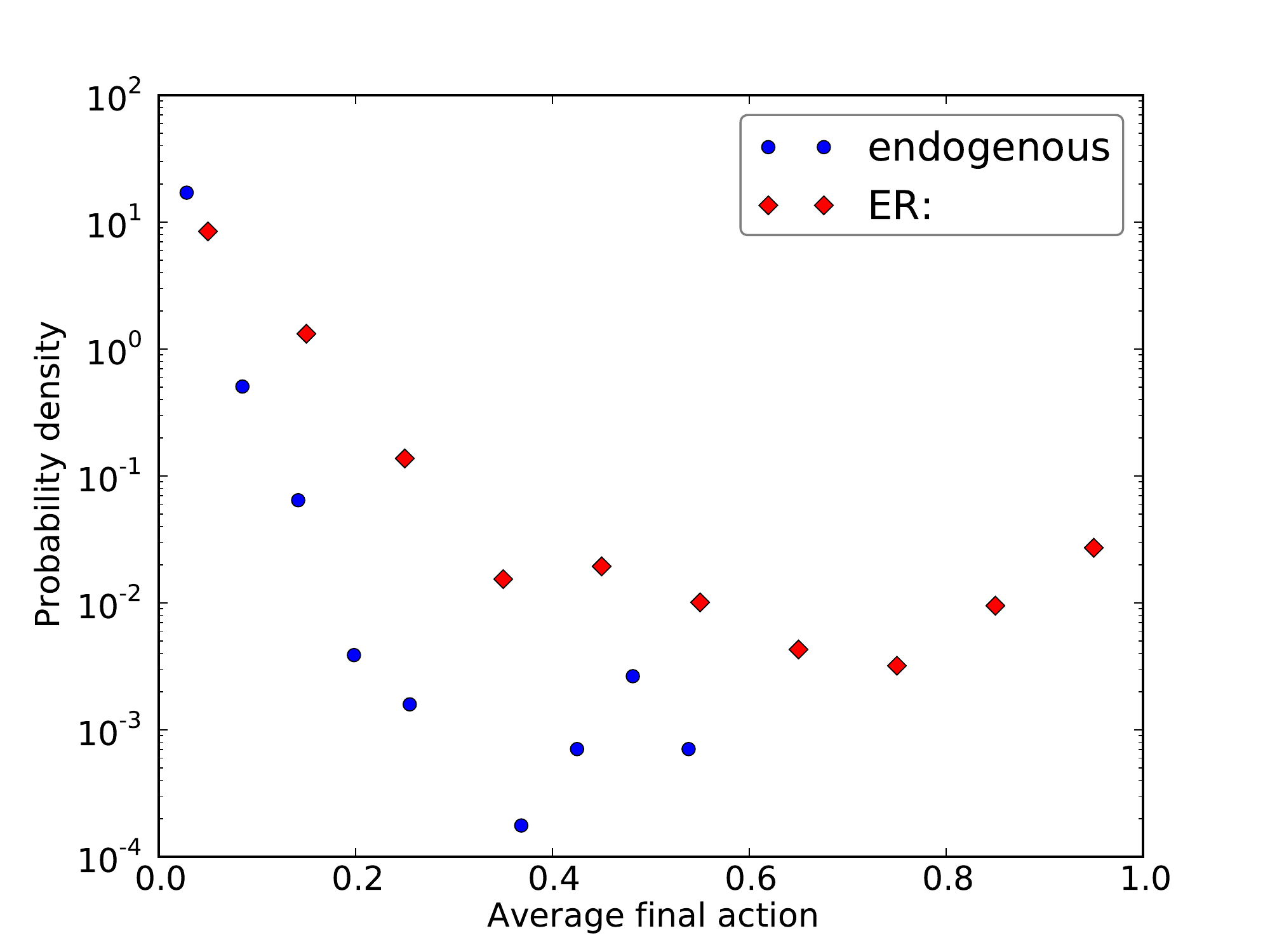}
\caption{Distribution of final action in ER networks vs. endogenous networks. This is without bias, i.e. the initial action is random based on the private belief only.}\label{FIG::ENDO_X_FINAL_DIST}
\end{figure}

\begin{figure}[ht]
\centering
\includegraphics[width=0.7\textwidth]{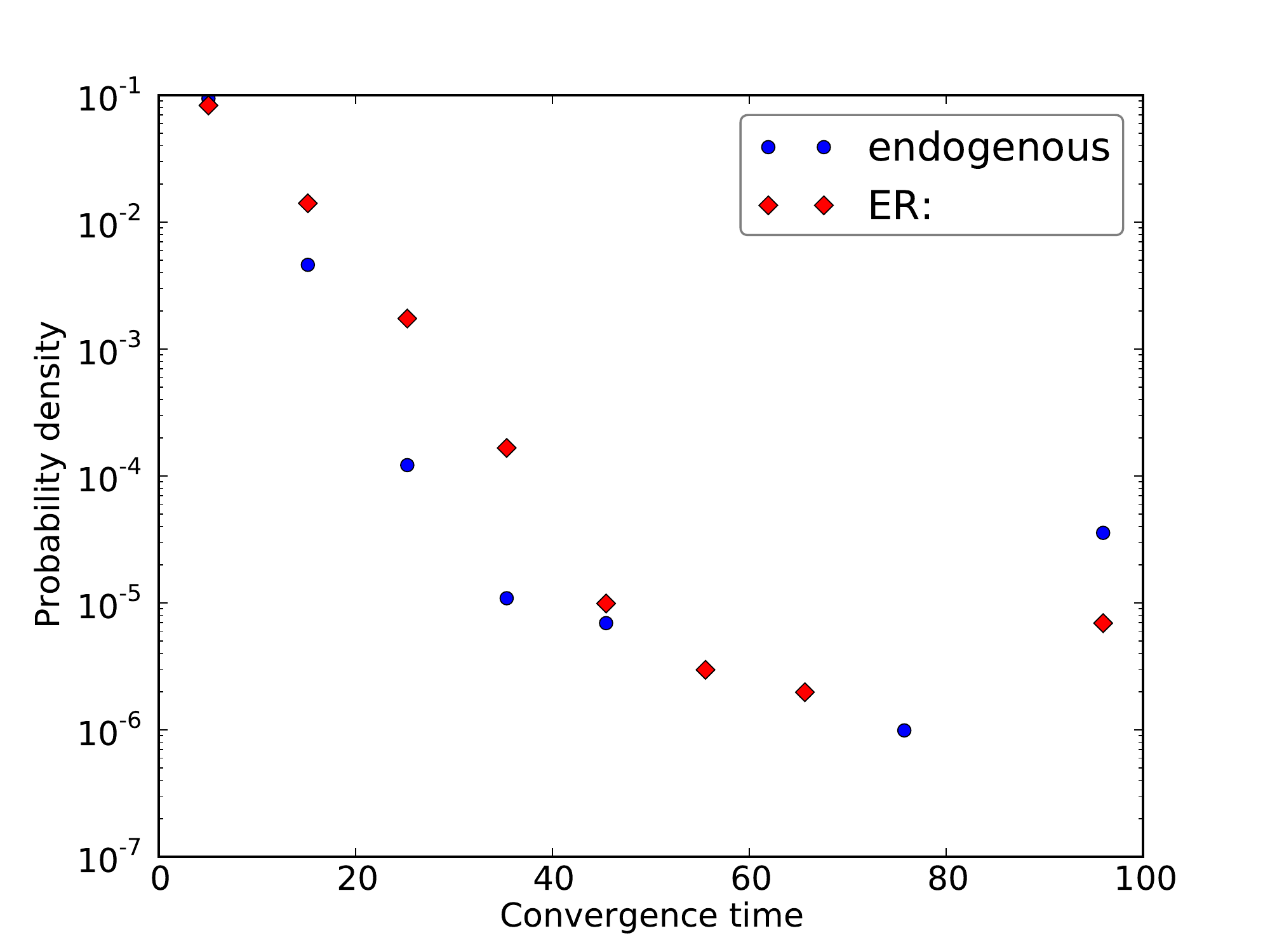}
\caption{Distribution of convergence time ER networks vs. endogenous networks. This is without bias, i.e. the initial action is random based on the private belief only.}\label{FIG::ENDO_CONV_T_DIST}
\end{figure}

\begin{figure}[ht]
\centering
\includegraphics[width=0.7\textwidth]{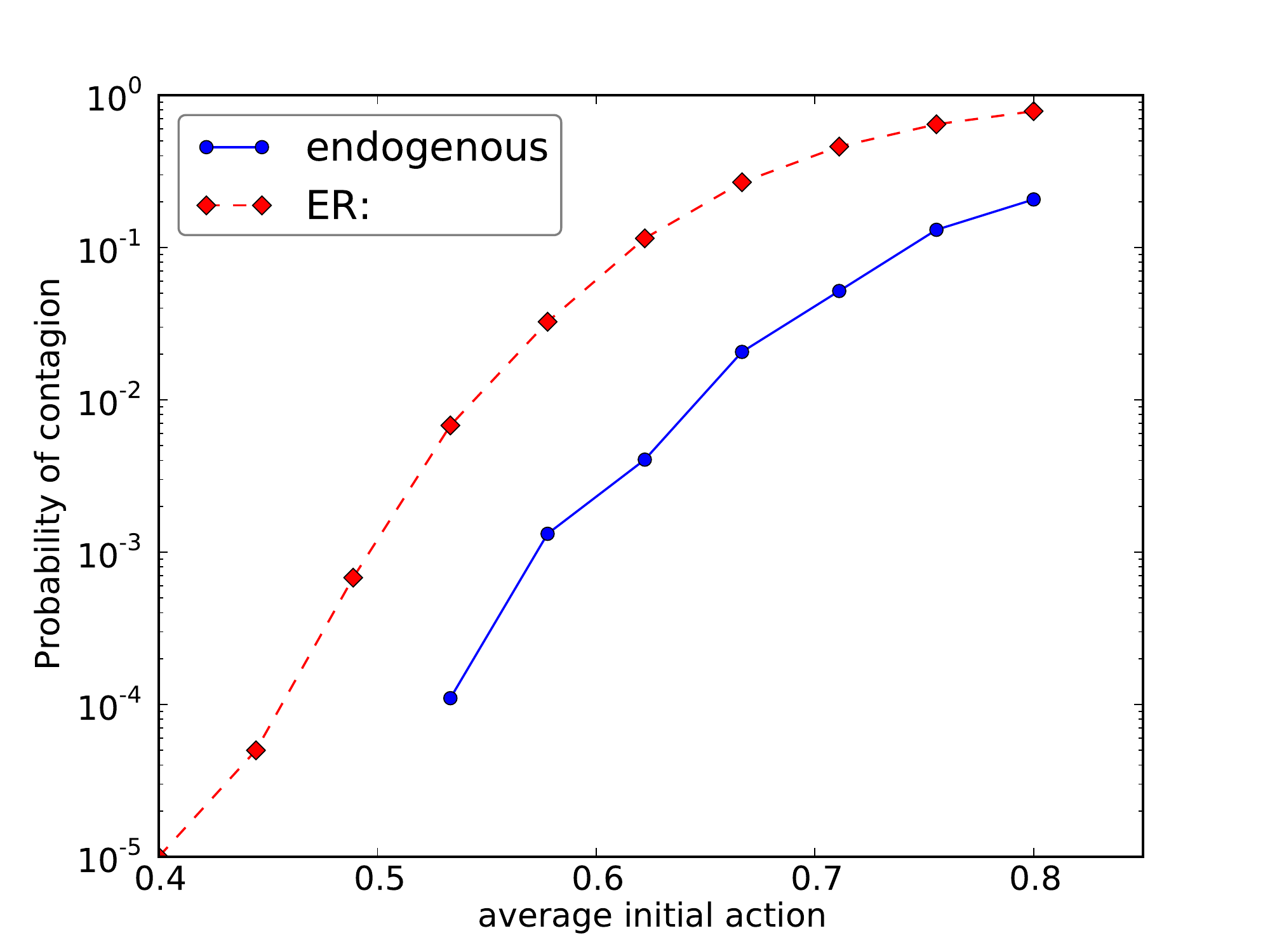}
\caption{Probability of contagion vs. initialization bias. Missing values correspond to zero frequency.}\label{FIG::ENDO_P_CONT_LOG}
\end{figure}

\clearpage
\section{Proofs}\label{Appendix::Proofs}

\begin{figure}[t]
\centering
\includegraphics[width=0.7\textwidth]{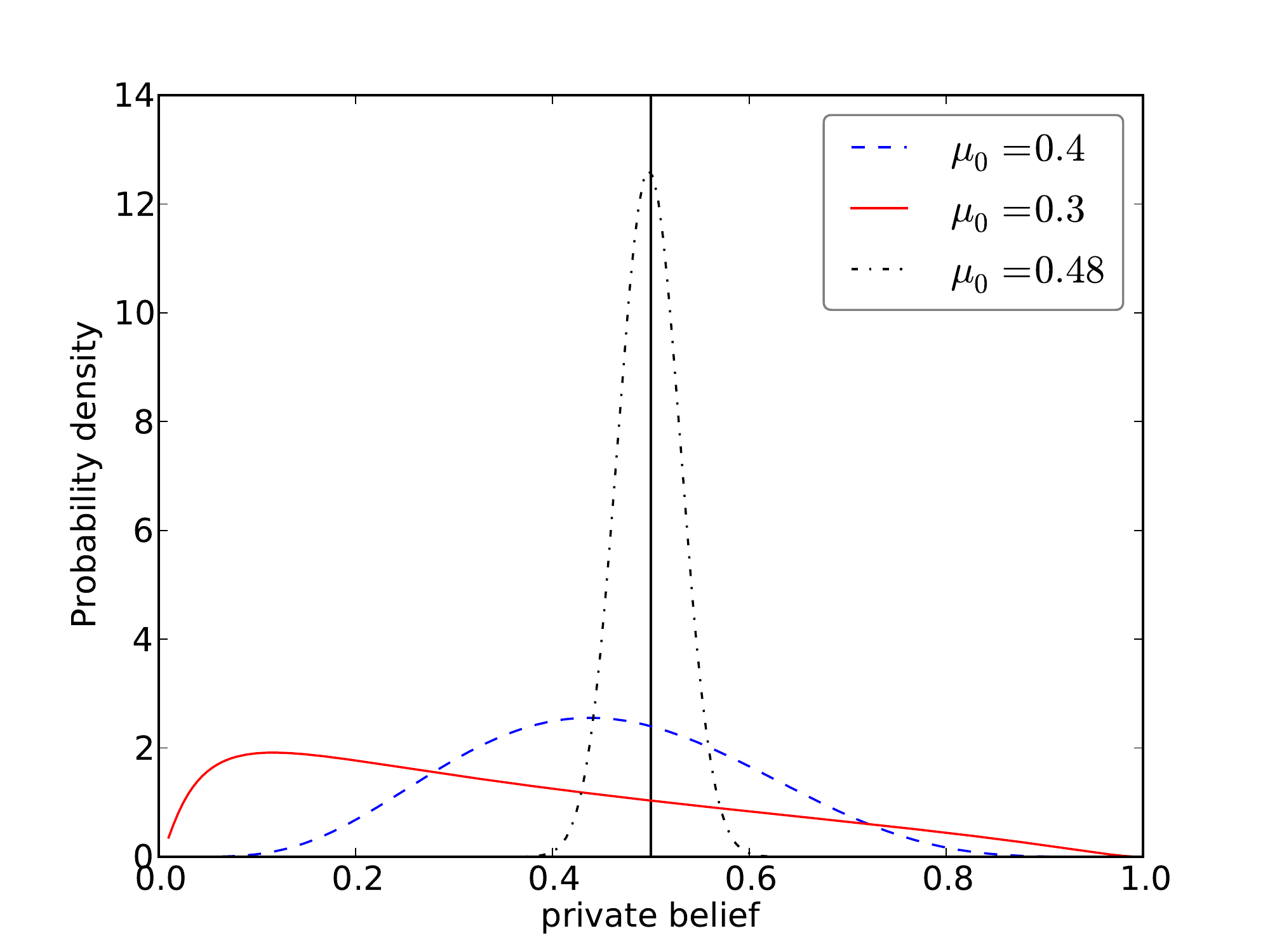}
\caption{Probability density function of the private belief given that the state of the world is $\theta=0$ $f_P(p \mid \theta = 0)$ for three values of $\mu_0 \in \left\lbrace 0.3,0.4,0.48 \right\rbrace$.}\label{Fig::AppPrivateBeliefDist}
\end{figure}

{\bf Proof of Proposition (\ref{Prop::Prop1})}. We use the notation $f_P(p^i \mid \theta = 0)$ to indicate that the functional form of the probability distribution of the private belief $p^i$ has be derived assuming that $\theta = 0$. Now, let $s(p^i)$ be the inverse of the private belief:
\begin{equation}
	s(p^i)=  \frac{\mu_i^2-\mu_1^2+2 \sigma^2 \log \left(\frac{1-p^i}{p^i}\right)}{2 (\mu_0-\mu_1)}
\end{equation}
The distribution of the private belief can be computed as follows:
\begin{equation}
	f_p(p^i) = \frac{\partial s(p^i) }{\partial p^i } f_s(s(p^i)).
\end{equation}
where the probability density function for signal $s$ is given as:
\begin{equation}
	f_s(s) = \frac{1}{\sqrt{2\pi}\sigma} \exp \left( \frac{-(s-\mu_0)^2}{2 \sigma^2} \right)
\end{equation}
and the private belief $p^i(s)$ is given by Equation (\ref{EQ:PrivateBelief}). Substituting in the expression for $s(p^i)$ and computing the partial derivative we obtain:
\begin{equation}
f_p(p^i \mid \theta = 0) = \frac{\left( \left(\frac{1-p^i}{p^i} - 1 \right) \sigma^2\right) \exp \left(-\frac{\left((\mu_0-\mu_1)^2-2 \sigma^2 \log \left(\frac{1}{p^i}-1\right)\right)^2}{8 \sigma^2 (\mu_0-\mu_1)^2}\right)}{\left(\sqrt{2 \pi } \sigma\right) ((1-p^i) (\mu_0-\mu_1))}
\end{equation}
Example distributions of the private belief are shown in Figure (\ref{Fig::AppPrivateBeliefDist}). We have $\mu_1 = 1 - \mu_0$ and $\sigma^2 = 0.1$. Note, that the majority of the probability density of the private belief is to the left of $0.5$ in all cases. Therefore, the private signal tends to produce private beliefs that yield the state matching action. If we increase the $\mid\mu_0 - 0.5\mid$ the distribution becomes more skewed towards the actual state of the world. Hence the private belief becomes more informative.

Now that we have defined the pdf of the private belief we can compute the probability that the agent chooses $x^i=0$ given some social belief $q=q^i$ as:
\begin{equation}
	\Pr(x^i = 0 \mid q^i = q) = \int_0^{1-q} f_p(p^i \mid \theta = 0) dp^i,
\end{equation}
where we use the notation $f_p(p^i \mid \theta = 0)$ to indicate that the functional form of $f_p$ has be derived assuming that $\theta = 0$. This result generalizes to all $\theta$ due to the symmetry of the signal structure. It can be shown that:
\begin{equation}
	\Pr(x^i = \theta \mid q^i = q) = \int_0^{1-q} f_p(p^i \mid \theta = 0) dp^i  =\int_{1-q}^1 f_p(p^i \mid \theta = 1) dp^i.
\end{equation}
Therefore we have derived an expression for the probability of choosing the correct action that does not depend on the actual state of the world but only on the signal structure and the social belief.

\end{appendix}

\end{document}